\documentclass[12pt]{article}
\input psfig.sty

\def\Re{{\cal R \mskip-4mu \lower.1ex \hbox{\it e}\,}}
\def\Im{{\cal I \mskip-5mu \lower.1ex \hbox{\it m}\,}}

\def\sub#1{_{\lower.25ex\hbox{$\scriptstyle#1$}}}
\def\tev{\,{\ifmmode\mathrm {TeV}\else TeV\fi}}
\def\gev{\,{\ifmmode\mathrm {GeV}\else GeV\fi}}
\def\mev{\,{\ifmmode\mathrm {MeV}\else MeV\fi}}
\def\mpl{\ifmmode \overline M_{Pl}\else $\overline M_{Pl}$\fi}
\def\to{\rightarrow}

\def\subw{_{\rm w}}
\def\mh{\ifmmode m\sbl H \else $m\sbl H$\fi}
\def\mch{\ifmmode m_{H^\pm} \else $m_{H^\pm}$\fi}
\def\mt{\ifmmode m_t\else $m_t$\fi}
\def\mc{\ifmmode m_c\else $m_c$\fi}
\def\mz{\ifmmode M_Z\else $M_Z$\fi}
\def\mw{\ifmmode M_W\else $M_W$\fi}
\def\mws{\ifmmode M_W^2 \else $M_W^2$\fi}
\def\mhs{\ifmmode m_H^2 \else $m_H^2$\fi}   
\def\mzs{\ifmmode M_Z^2 \else $M_Z^2$\fi}
\def\mts{\ifmmode m_t^2 \else $m_t^2$\fi}
\def\mcs{\ifmmode m_c^2 \else $m_c^2$\fi}
\def\mchs{\ifmmode m_{H^\pm}^2 \else $m_{H^\pm}^2$\fi}
\def\ztwo{\ifmmode Z_2\else $Z_2$\fi}
\def\zone{\ifmmode Z_1\else $Z_1$\fi}
\def\mtwo{\ifmmode M_2\else $M_2$\fi}
\def\mone{\ifmmode M_1\else $M_1$\fi}
\def\tb{\ifmmode \tan\beta \else $\tan\beta$\fi}
\def\xw{\ifmmode x\subw\else $x\subw$\fi}
\def\ch{\ifmmode H^\pm \else $H^\pm$\fi}
\def\lum{\ifmmode {\cal L}\else ${\cal L}$\fi}
\def\inpb{\,{\ifmmode {\mathrm {pb}}^{-1}\else ${\mathrm {pb}}^{-1}$\fi}}
\def\infb{\,{\ifmmode {\mathrm {fb}}^{-1}\else ${\mathrm {fb}}^{-1}$\fi}}
\def\epem{\ifmmode e^+e^-\else $e^+e^-$\fi}
\def\ppb{\ifmmode \bar pp\else $\bar pp$\fi}
\def\bsg{\ifmmode B\to X_s\gamma\else $B\to X_s\gamma$\fi}
\def\bsll{\ifmmode B\to X_s\ell^+\ell^-\else $B\to X_s\ell^+\ell^-$\fi}
\def\bstt{\ifmmode B\to X_s\tau^+\tau^-\else $B\to X_s\tau^+\tau^-$\fi}
\def\lamt{\ifmmode \tilde\lambda\else $\tilde\lambda$\fi}
\def\shat{\ifmmode \hat s\else $\hat s$\fi}
\def\that{\ifmmode \hat t\else $\hat t$\fi}
\def\uhat{\ifmmode \hat u\else $\hat u$\fi}

\newskip\zatskip \zatskip=0pt plus0pt minus0pt
\def\matth{\mathsurround=0pt}
\def\lsim{\mathrel{\mathpalette\atversim<}}
\def\gsim{\mathrel{\mathpalette\atversim>}}
\def\atversim#1#2{\lower0.7ex\vbox{\baselineskip\zatskip\lineskip\zatskip
  \lineskiplimit 0pt\ialign{$\matth#1\hfil##\hfil$\crcr#2\crcr\sim\crcr}}}

\renewcommand{\thefootnote}{\fnsymbol{footnote}}

\begin{document} \begin{titlepage} 
\rightline{\vbox{\halign{&#\hfil\cr
&SLAC-PUB-9146\cr
%&March 8, 2002\cr
}}}
\begin{center}

{\Large\bf Precision Measurements and Fermion Geography in the Randall-Sundrum 
Model Revisited}
\footnote{Work supported by the Department of 
Energy, Contract DE-AC03-76SF00515}
\medskip

\normalsize 
{\bf \large J.L. Hewett, F.J. Petriello and T.G. Rizzo}
\vskip .2cm
Stanford Linear Accelerator Center \\
Stanford University \\
Stanford CA 94309, USA\\
\vskip .2cm

\end{center} 

\begin{abstract} 

We re-examine the implications of allowing fermion fields to 
propagate in the five-dimensional bulk of the Randall-Sundrum (RS)
localized gravity model.  We find that mixing between the Standard 
Model top quark and its Kaluza Klein excitations generates large
contributions to the $\rho$ parameter and consequently
restricts the fundamental RS 
scale to lie above 100 TeV.  To circumvent this bound we 
propose a `mixed' scenario which localizes the third generation fermions on 
the TeV brane and allows the lighter generations to propagate in the full 
five-dimensional bulk.  We show that this construction naturally reproduces 
the observed $m_c / m_t$ and $m_s / m_b$ hierarchies.  We explore the 
signatures of this scenario in precision measurements and future high energy 
collider experiments.  We 
find that the region of parameter space that addresses the hierarchies of 
fermion Yukawa couplings permits a Higgs boson with a mass of 500 GeV
and remains otherwise invisible at the LHC.  However, the entire parameter 
region consistent with electroweak precision data is testable at future 
linear colliders.  We briefly discuss possible constraints on this scenario 
arising from flavor changing neutral currents.

\end{abstract} 

%\vskip0.45in
%\begin{center}

%Submitted to Physical Review {\bf D}.

%\end{center}

\renewcommand{\thefootnote}{\arabic{footnote}} \end{titlepage}

%%%%%%%%%%%%%%%%%%%%%%%%%%%%%%%---- Put text here

\section{Introduction} 

The Randall-Sundrum (RS) model~\cite{Randall:1999ee} offers a new approach to 
the hierarchy problem.  This scheme proposes that our four-dimensional world 
is embedded in a five-dimensional spacetime described by the metric
\begin{equation}
ds^2=e^{-2 \sigma (y)}\eta_{\mu\nu}dx^\mu dx^\nu -dy^2\,,
\end{equation}
where $\sigma (y) = k |y|$, and with the $5^{th}$ dimensional 
coordinate $y=r_c\phi$ being compactified 
on an $S^1/Z_2$ orbifold bounded
by branes of opposite tension at the fixed points $y=0$ (known as 
the Planck brane) and 
$y=\pi r_c$ (TeV-brane).  The parameter $k$ describes the curvature of the 
space (with the five-dimensional curvature invariant being given by ${\cal R}
=-20k^2$) and is of order the five-dimensional Planck scale, $M_5$, so 
that no additional hierarchy exists. 
Self-consistency of the classical theory requires~\cite{Randall:1999ee} that 
$|{\cal R}| \leq M_5^2$ so that quantum gravitational effects can be 
neglected. The space between the two branes is $AdS_5$ and their separation, 
$\pi r_c$, can be naturally stabilized~\cite{stable} with 
$kr_c \simeq 11-12$; we employ $kr_c=11.27$ in the numerical results that 
follow.  For such values of $kr_c$ any mass of order the Planck scale on the 
$y=0$ brane appears 
to be suppressed by an amount $e^{-\pi kr_c}\sim 10^{-15}$ on the TeV brane.
The presence of the exponential warp factor $e^{-\sigma(y)}$ thus
naturally generates the hierarchy between the Planck and electroweak (EW)
scales.  The scale of physics on the TeV-brane is given by $\Lambda_\pi = 
\mpl e^{-kr_c\pi}\sim TeV$, where $\mpl$ is the
reduced Planck scale.
Integration over the extra dimension of the five dimensional 
RS action yields the relationship between the 4-dimensional Planck 
scale  and the scales $k$ and $M_5$:
\begin{equation}
\mpl^2=M_5^3/k\,.
\end{equation}
Together, this relation and the inequality $|{\cal R}|\leq M_5^2$ 
imply that the ratio 
$k/\mpl$ cannot be too large, and suggests that $k/\mpl \lsim 0.1-1$. 

In the original RS framework, gravity propagates freely throughout the
bulk while the Standard Model (SM) fields are constrained to the TeV-brane.
The graviton KK states have non-trivial wave 
functions in the extra dimension due to the warp factor, 
and have masses given by 
$m_n=x_nke^{-\pi kr_c}$, where the $x_n$ are the unequally spaced 
roots of the Bessel function $J_1$ and $n$ labels the KK excitation level. 
The first graviton excitation thus naturally has a mass of order a TeV.
The $n>0$ KK states couple to fields on the TeV-brane with a strength
of $\Lambda_\pi^{-1}$.   The graviton KK states can thus
be produced in colliders as TeV-scale resonances with TeV$^{-1}$-size 
couplings to matter~\cite{Davoudiasl:1999jd}. 

For additional freedom in model building,
the original RS model has been extended to allow various subsets of the SM 
fields to reside in the bulk in the limit that the 
back-reaction on the metric can be neglected.  This possibility allows
for new techniques to address gauge coupling unification, supersymmetry
breaking, the neutrino mass spectrum, and the fermion mass hierarchy.
Placing the gauge fields of the SM alone in the bulk is 
problematic~\cite{Davoudiasl:1999tf,Pomarol:1999ad,Csaki:2002gy}, 
as all of the gauge KK excitations then have large couplings to the 
remaining fields on 
the TeV-brane; these couplings take on the value
$\sqrt {2\pi kr_c}\, g\simeq 8.4\, g$, where $g$ is the corresponding
SM gauge coupling.  EW precision data then constrain
the masses of the first KK gauge states to be 
in excess of 25-30 TeV, thus requiring $\Lambda_\pi$ to be in excess of 
100 TeV~\cite{Davoudiasl:1999tf}.  This introduces a new hierarchy between 
$\Lambda_{\pi}$ and the
EW scale, and therefore this scenario is highly disfavored.  It was
subsequently shown that these constraints can be softened
by also placing the SM fermions in the bulk \cite{Grossman:1999ra,big} and 
giving them a common five-dimensional mass $m=k\nu$.
This leads to further model building 
possibilities provided $\nu$ is in the range $-0.8 \lsim \nu \lsim 
-0.3$~\cite{Davoudiasl:2000wi}; for larger values of $\nu$ the former strong 
coupling 
regime is again entered, while for smaller values potential problems with 
perturbation theory can arise~\cite{Davoudiasl:2000my}.  In the absence of 
fine-tuning, these scenarios require that 
the Higgs field which breaks the symmetry of the SM remains on the TeV brane,  
as when bulk Higgs fields are employed, the experimentally observed pattern of 
$W$ and $Z$ masses cannot be reproduced.  In this case, the $W$ and $Z$ 
obtain a common KK mass in addition to the usual contribution from 
the Higgs vev. It is then impossible to simultaneously maintain the two 
tree-level SM relationships $M_Z \cos \theta_w=M_W$ and 
$e=g \sin \theta_w$~\cite{Pomarol:1999ad,big,Davoudiasl:2000wi}. 

In this paper we re-examine the possibility of allowing
the SM fermions to propagate 
in the RS bulk.  We show that if the third generation of fermions resides 
in the bulk, then large mixing between the fermion zero modes and their
KK tower states is induced by the SM Higgs vev and the large 
top Yukawa coupling.  This mixing results in contributions to
$\delta \rho$ or $T$~\cite{Peskin:1990zt} which greatly exceeds the 
bound set by current precision EW measurements~\cite{pm}. The only way to 
circumvent this problem is to raise the mass of the first KK gauge state 
above 25 TeV for any value of $\nu$ in its viable range, which again implies
a higher value of $\Lambda_\pi$. 
Unless we are willing to fine tune $\Lambda_\pi$, we must then require 
the third generation fields to remain on the TeV-brane so that they have no KK 
excitations.  If we treat the three generations symmetrically we must 
localize all of the fermions on the TeV-brane, and also confine the SM gauge 
sector to the TeV brane as discussed in the previous paragraph.  We instead
propose here a `mixed' scenario which places the first two generations 
of fermions in the 
bulk and localizes the third on the wall.  We find that mixing of the KK 
towers of these 
lighter generations with their zero modes does not yield a dangerously large 
value of $\delta \rho$ provided that $\nu \gsim -0.6$.
Furthermore, we show that values of $\nu$ 
near $-0.4$ to $-0.5$ may help explain the mass hierarchies 
$m_c/m_t$ and $m_s/m_b$.  
We explore the possible signatures of this scenario at the LHC and
future linear colliders,as well as 
in precision measurements and flavor changing 
neutral currents (FCNC); we find that the same parameter space which addresses 
the fermion mass hierarchies also allows a Higgs boson with a mass of 500 GeV,
and is otherwise invisible at the LHC.

The outline of this paper is as follows.  In Section II we give a brief 
overview of the mechanics of the RS model which we will need for subsequent 
calculations. In Section III we examine the contributions to the $\rho$ 
parameter when the third generation is in the bulk and show that this scenario 
is highly disfavored.  We examine the present bounds on the KK mass spectrum 
in our mixed scenario that arise from precision EW data 
in Section IV.  We demonstrate that SM Higgs masses as large as 
500 GeV are now allowed by the electroweak fit since these contributions can 
be partially ameliorated from those of the KK states. Section V explores the 
implications of this scenario for the LHC, while Section VI examines the 
signatures at a future \epem\ linear collider 
and at GigaZ. In particular, we show 
that the KK states in this model lie outside the kinematically limited range 
of the LHC but yield observable indirect effects at a linear collider.
Finally, we discuss constraints from
FCNC in Section VII, and present our conclusions in Section VIII.

\section{The Standard Model Off the Wall}

We present here a cursory formulation of the RS model in the case where
the SM gauge
and fermion fields propagate in the bulk; we refer the reader  
to~\cite{Davoudiasl:2000wi} for a thorough introduction.  

We begin by considering a $SU$(N) gauge theory defined by the action
\begin{equation}
S_A = - \frac{1}{4} \int d^{5}x \sqrt{-G} \, G^{MK}G^{NL}F_{KL}F_{MN} \,\, ,
\end{equation}
where $G^{\alpha \beta}=e^{-2\sigma} \left( \eta^{\alpha \beta} +\kappa_{5}
h^{\alpha \beta} \right)$,
$\sqrt{-G} \equiv |det(G_{MN})|^{1/2} = e^{-4 \sigma}$, 
with $\kappa_5 = 2M_{5}^{-3/2}$, $\eta_{\alpha\beta}$ being the 
Minkowski metric
with signature -2, and $h_{\alpha\beta}$ represents the graviton
fluctuations.  $F_{MN}$ is
the 5-dimensional field strength tensor given by
\begin{equation}
F_{MN} = \partial_M A_N -\partial_N A_M +ig_5 \, [A_M,A_N] \,\, ,
\end{equation}
and $A_M$ is the matrix valued 5-dimensional gauge field and $g_5$ is
the corresponding 5-dimensional gauge coupling.  We impose the 
gauge condition $A_5 = 0$; this is consistent with 5-dimensional gauge
invariance~\cite{Davoudiasl:1999tf}, and with the $Z_2$-odd parity
assigned to $A_5$ 
to remove its zero mode from the TeV-brane action.  To derive the
effective 4-dimensional theory we expand $A_{\mu}$ as
\begin{equation}
A_{\mu}(x,\phi) = \sum_{n=0}^{\infty} A_{\mu}^{(n)} \frac{\chi^{(n)}
(\phi)}{\sqrt{r_c}} \,\, ,
\label{gaugeexpand}
\end{equation}
and require that the bulk wavefunctions $\chi^{(n)}$
satisfy the orthonormality constraint
\begin{equation}
\int_{-\pi}^{\pi} d\phi \, \chi^{(m)} \chi^{(n)} = \delta^{mn} \,\, .
\label{orthogauge}
\end{equation}
We obtain a tower of massive KK gauge fields $A^{(n)}_{\mu}$, with $n \geq 1$,
and a massless zero mode $A^{(0)}_{\mu}$.  The KK masses $m^{A}_n$ are 
determined by the eigenvalue equation
\begin{equation}
-\frac{1}{r_{c}^{2}} \frac{d}{d\phi} \left( e^{-2\sigma} \frac{d}{d\phi}
\chi^{(n)} \right) = \left( m_{n}^A\right)^2 \chi^{(n)} \,\, .
\end{equation}
This yields $m^{A}_{n}=x^{A}_{n} k e^{-kr_c \pi}$ on the TeV-brane,
where the $x_n^A$ are given in
\cite{Davoudiasl:1999tf}, with the first few
numerical values being given by $x^{A}_{1} \simeq 2.45$, 
$x^{A}_{2} \simeq 5.57$, $x^{A}_{3} \simeq 8.70$. 
Explicit expressions for the bulk wavefunctions 
$\chi^{(n)}$ also contain the first order Bessel functions $J_1$ and $Y_1$, 
and can be found in~\cite{Davoudiasl:1999tf,Davoudiasl:2000wi}; we note here 
only that the zero mode wavefunction is $\phi$ independent with $\chi^{(0)} = 
1/ \sqrt{2\pi}$.  

We now add a fermion field charged under this gauge group, and able to
propagate in the bulk.  The action for this field is 
\begin{equation}
S_F = \int d^{4}x \int dy \sqrt{-G} \left[V^{M}_{n} \left( \frac{i}{2}
\bar{\Psi}\gamma^{n}\stackrel{\leftrightarrow} {D}_{M}
\Psi +{\rm h.c.} \right) -{\rm sgn}(y) \,
m \bar{\Psi}\Psi \right] \,\, ,
\label{fermaction}
\end{equation}
where h.c. denotes the hermitian conjugate, $V^{M}_{\mu}=e^{\sigma}
\delta^{M}_{\mu}$, $V^{5}_{5}=-1$, $\gamma^{n}=(\gamma^{\mu},i\gamma_{5})$,  
$D_{M}$ is the covariant derivative, and $m$ is the 5-dimensional Dirac mass 
parameter.  This 5-dimensional fermion is necessarily vector-like;
we wish to obtain a chiral zero mode from its KK expansion.
We follow~\cite{Grossman:1999ra} and expand the chiral components of the
5-dimensional field as
\begin{equation}
\Psi_{L,R}(x,\phi) = \sum_{n=0}^{\infty} \psi^{(n)}_{L,R}(x) \frac{e^{2\sigma}}
{\sqrt{r_c}} f^{(n)}_{L,R} (\phi) \,\, ,
\label{fermexpand}
\end{equation}
and require the orthonormality conditions
\begin{equation}
\int_{-\pi}^{\pi} e^{\sigma} f^{(m)*}_{L}f^{(n)}_{L} = 
\int_{-\pi}^{\pi} e^{\sigma} f^{(m)*}_{R}f^{(n)}_{R} =\delta^{mn} \,\, .
\label{orthoferm}
\end{equation}
The $Z_2$ symmetry of the 5-dimensional mass term in the action
forces $f^{(n)}_{L}$ and
$f^{(n)}_{R}$ to have opposite $Z_2$ parity; we choose $f^{(n)}_{L}$ to be
$Z_2$ even and $f^{(n)}_{R}$ to be $Z_2$ odd.  As shown 
in~\cite{Grossman:1999ra}, this removes $f^{(0)}_{R}$ from the TeV-brane action,
and we obtain the chiral zero mode $f^{(0)}_{L}$ necessary for 
construction of the SM.  The KK states form a tower of massive vector 
fermions.  The zero mode wavefunction is 
\begin{equation}
f^{(0)}_{L} =\frac{e^{\nu\sigma}}{N^{L}_{0}} \,\, ,
\label{fermzero}
\end{equation}
where $\nu = m/k$ and is expected to be of order unity,
and $N^{L}_{0}$ is determined from the orthonormality 
constraint of Eq.~\ref{orthoferm}.  Explicit expressions for the KK fermion 
masses and wavefunctions are given in~\cite{Davoudiasl:2000wi}; we note here
that $m_{n}^{F}=m_{n}^{A}$ when
$\nu = -0.5$, and that $m_{n}^{F}>m_{n}^{A}$ for all other values of $\nu$.

\begin{figure}[htbp]
\centerline{
\hspace*{-0.8cm}
\psfig{figure=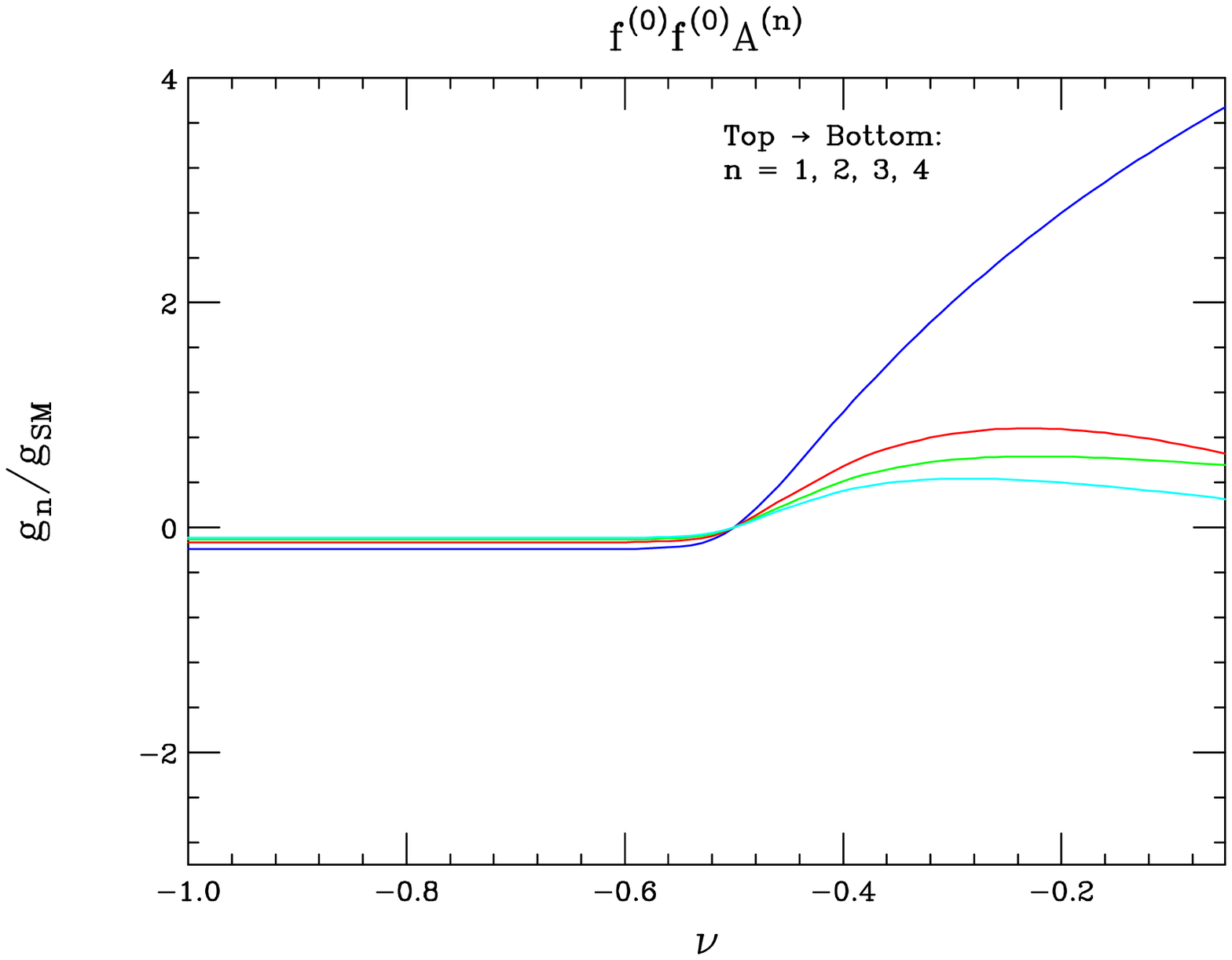,height=7.2cm,width=8.5cm,angle=0}
\hspace*{0.2cm}
\psfig{figure=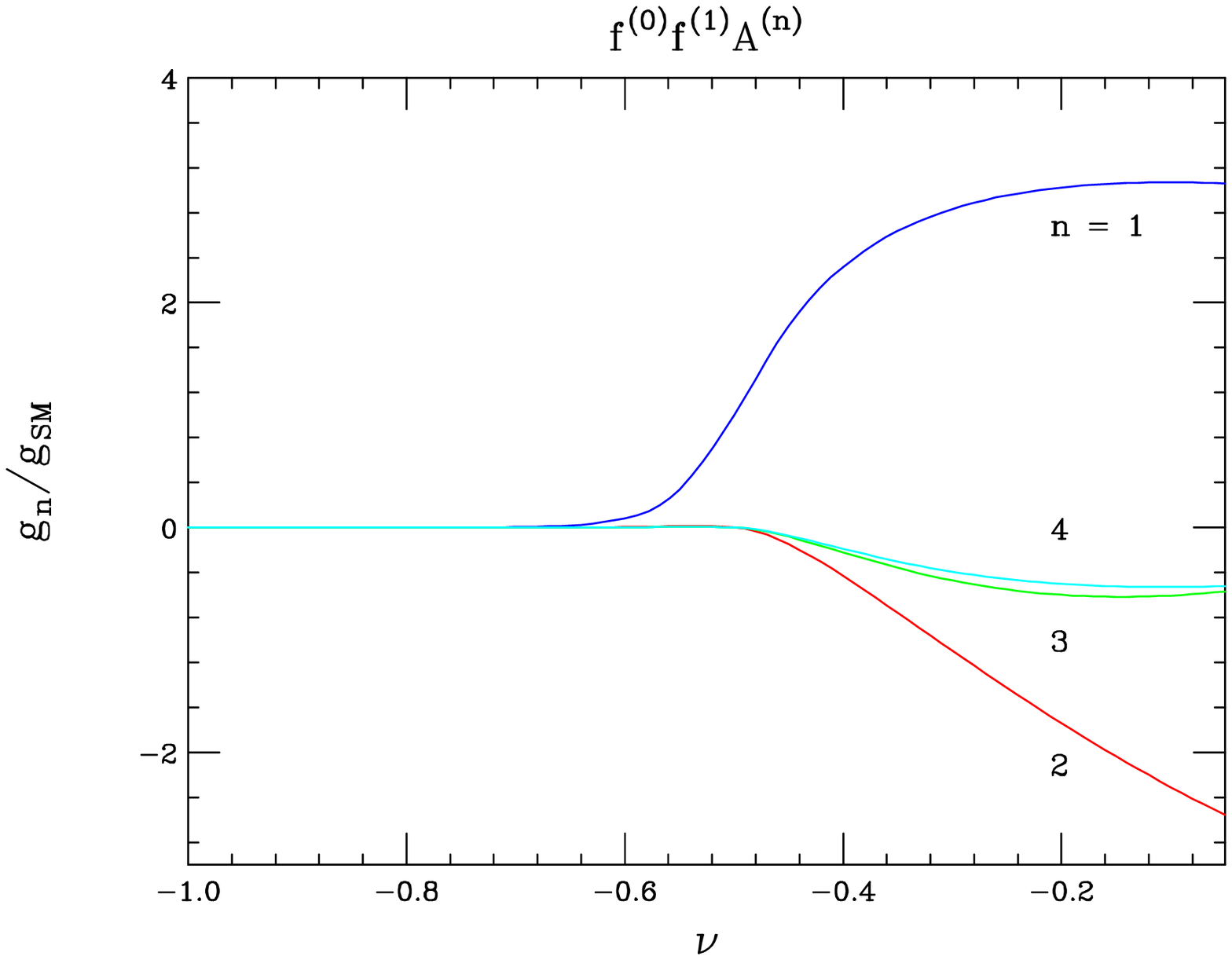,height=7.2cm,width=8.5cm,angle=0}}
\caption{The coefficients $C^{00n}_{f\bar{f}A}$ (left) and 
$C^{01n}_{f\bar{f}A}$ (right) for $n=1,\ldots,4$ as functions of the fermion
bulk mass parameter $\nu$.}
\label{FAcoups}
\end{figure}

Inserting the KK expansions of both the gauge and fermion fields into the
covariant derivative term in Eq.~\ref{fermaction}, we find that the ratios of
fermion-gauge KK couplings to the corresponding 4-dimensional coupling are
\begin{equation}
C^{mnq}_{f\bar{f}A} = \sqrt{2\pi}\int_{-\pi}^{\pi} 
d\phi \, e^{\sigma} f^{(m)}_{L} f^{(n)}_{L} \chi^{(q)} \,\, ,
\label{FAcoef}
\end{equation}
where $m, n, q$ label the excitation state.
The coefficients $C^{00n}_{f\bar{f}A}$ and $C^{01n}_{f\bar{f}A}$ are shown
in Fig.~\ref{FAcoups} for $n=1,\ldots,4$ as functions of $\nu$.  Notice that 
$C^{001}_{f\bar{f}A}$ vanishes at $\nu = -0.5$ and remains small for
$\nu < -0.5$; this fact will be crucial in our later analysis.

In addition, the ratios of the KK triple gauge couplings (TGCs) to the 
TGC of the 4-dimensional theory are given by
\begin{equation}
C^{mnq}_{AAA} = \frac{g^{(mnq)}}{g} = \sqrt{2\pi} \int_{-\pi}^{\pi} 
d\phi \, \chi^{(m)} \chi^{(n)} \chi^{(q)} \,\, ,
\end{equation}
where we have identified $g=g_5 /\sqrt{2\pi r_c}$.  Using the zero mode 
wavefunction $\chi^{(0)} = 1/ \sqrt{2\pi}$ and the orthonormality constraint 
of Eq.~\ref{orthogauge}, we find that $C^{n00}_{AAA}=0$ when $n > 0$; no 
coupling exists between two zero mode gauge particles and a KK gauge state.

%\vspace*{-0.1cm}
\noindent

We will also require the couplings between gauge KK states and the fermion 
fields which are localized on the TeV-brane \cite{Davoudiasl:1999tf}.  
The relevant action is
\begin{equation}
S_F = \int d^{4}x \int d\phi \sqrt{-G} \left[V^{M}_{n} \left( \frac{i}{2}
\bar{\psi}\gamma^{n}\stackrel{\leftrightarrow} {D}_{M}
\psi +{\rm h.c.} \right) \right] \delta 
\left(\phi-\pi \right) \,\, .
\end{equation}
Inserting the expansion of Eq.~\ref{gaugeexpand} into this expression, 
letting $\psi \rightarrow e^{3 \sigma/2} \psi$, and setting $g = g_5 /
\sqrt{2\pi r_c}$, we find that the ratio of the n$^{th}$ KK gauge coupling
to localized fermions relative to the corresponding SM coupling is
\begin{equation}
C^{n}_{f\bar{f}A} = \frac{\chi^{(n)}(\pi)}{\chi^{(0)}(\pi)} \,\, .
\label{wallcoup1}
\end{equation}
Utilizing the approximate expressions for the KK gauge wavefunctions 
in~\cite{Davoudiasl:2000wi}, these become
\begin{equation}
C^{n}_{f\bar{f}A} \approx \left( -1 \right)^{n+1} \sqrt{2\pi kr_c} \,\, .
\label{wallcoup2}
\end{equation}

We now consider the final ingredient required for construction of the SM,
the Higgs boson.  As discussed in the introduction, the Higgs field
must be confined to the TeV-brane to correctly break the electroweak symmetry.
Its action can therefore be expressed as
\begin{equation}
S_H = \int d^{4}x \int dy \sqrt{-G} \left\{ G^{MN}D_{M}H \left( D_{N}H 
\right)^{\dagger} - V(H) \right\} \delta(y-r_c \pi)\,\, ,
\end{equation}
where $V(H)$ is the Higgs potential and $D_M$ the covariant derivative.  
To properly normalize the Higgs field kinetic term we must rescale
$H \rightarrow e^{\sigma} H$; we then expand $H$ around its vev, $v$, 
insert the expansion of Eq.~\ref{gaugeexpand} into the covariant derivative, 
and identify $g = g_5 /\sqrt{2\pi r_c}$.  We find the gauge field mass terms
\begin{equation}
S_{H,mass} = \frac{1}{2} \sum_{m,n=0}^{\infty} a_{mn} \int d^{4}x \, m_{A,0}^2 
\, A^{(m)}_{\mu} A^{(n),\mu} \,\, ,
\label{genmass}
\end{equation}
where $m_{A,0}$ is the gauge field mass of the 4-dimensional theory
corresponding to the zero-mode of the gauge KK tower, and 
\begin{equation}
a_{mn} = 2\pi \chi^{(m)}(\pi)\chi^{(n)}(\pi) \,\, .
\label{mixterms}
\end{equation}
We must diagonalize the full mass matrix, including the contributions arising 
from the KK reduction, to obtain the physical spectrum;
we will do so for the SM gauge fields in a later section.

We now examine the mixing between fermion KK states induced by the Higgs 
field.  When fermion fields are confined to the TeV-brane, no such
mixing occurs; we therefore consider only the case where the fermions
propagate in the bulk.  The coupling between the Higgs and KK fermions is
\begin{equation}
S_{f\bar{f}H}= \frac{\lambda^{'}}{k} \int d^{4}x \int dy \sqrt{-G} \left\{ 
H^{\dagger}
\Psi_{D} \Psi^{c}_{S} + {\rm h.c.} \right\} \delta \left( y-r_c \pi \right) 
\,\, ,
\end{equation}
where $\lambda^{'}$ is the 5-dimensional Yukawa coupling, and $k$ has been
introduced to make $\lambda^{'}$ dimensionless.  Both $\Psi_D$ and 
$\Psi^{c}_{S}$ are left-handed Weyl fermions; we have introduced 
the subscripts $D$ and $S$ for these fields to indicate that in the SM,
the Higgs couples $SU(2)_L$ doublets to $SU(2)_L$ singlets.  After
diagonalization of the mass matrix, the hermitian conjugates of the singlet
fields will combine with the appropriate doublets to form Dirac fermions.  
Since, for our analysis, we are interested only in the contributions to the 
fermion masses arising from this action, 
we again rescale the Higgs field by $e^{\sigma}$, set the Higgs field equal 
to its vev and expand the left-handed
fermion wavefunctions as in Eq.~\ref{fermexpand}.  Identifying 
\begin{equation}
\lambda = \frac{\lambda^{'} \left(f^{(0)}_{L}(\pi) \right)^2 e^{kr_c \pi}}{
kr_c} 
\label{yukawa}
\end{equation}
as the 4-dimensional Yukawa coupling, we find the fermion mass terms
\begin{equation}
S_{f,mass} = \sum_{m,n=0}^{\infty} b_{mn} \int d^{4}x \left\{ m_{f,0} 
\psi^{(m)}_{D} 
\psi^{c,(n)}_{S} +{\rm h.c.} \right\}\,\, ,
\end{equation}
where $m_{f,0}$ is the zero mode mass obtained when the KK states decouple, and
\begin{equation}
b_{mn}=\frac{f^{(m)}_{L}(\pi) f^{(n)}_{L}(\pi)}{\left(f^{(0)}_{L}(\pi) 
\right)^2} \,\, .
\end{equation}
Since $f^{(n)}_{L}(\pi)$ is approximately the 
same function for all $n \geq 1$, 
we will set $b_{0n} =\sqrt{f}$ and $b_{mn}=f$, with $m \neq n \neq 0$, 
in our analysis, where $f$ is a 
$\nu$ dependent quantity that measures the strength of the mixing. 
This parameter is explicitly given by
\begin{equation}
f= 2 {1-e^{-k\pi r_c(1+2\nu)}\over 1+2\nu}\,.
\end{equation}
These mass terms  must
be diagonalized in conjunction with the contributions from the
KK reduction of Eq.~\ref{fermaction}; we will do so for the SM $b$ 
and $t$ quarks in the next section.

\begin{figure}[htbp]
\centerline{
\psfig{figure=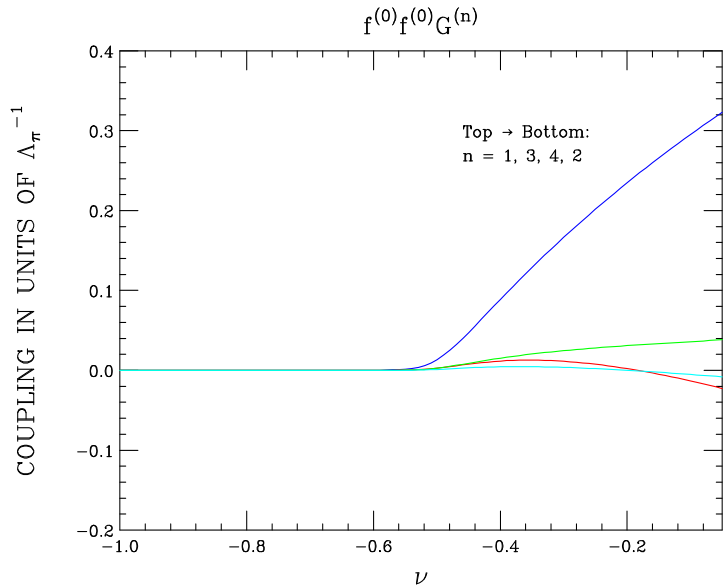,height=7.7cm,width=9.5cm,angle=0}}
\caption{The coupling strengths $C^{00n}_{f\bar{f}G}$ for $n=1,\ldots,4$ as
functions of the fermion bulk mass parameter $\nu$ in units of 
$\Lambda_{\pi}$. }
\label{FGcoups}
\end{figure}

To complete our discussion of the RS model, we must briefly discuss the 
KK gravitons it contains.  We parameterize the 5-dimensional metric as
\begin{equation}
G_{\alpha\beta}=e^{-2\sigma}\left(\eta_{\alpha\beta} +\kappa_{5} 
h_{\alpha\beta} \right) \,\, ,
\end{equation}
where $\kappa_5 = 2M_{5}^{-3/2}$, $\eta_{\alpha\beta}$ is the Minkowski metric
with signature -2, and the fluctuations of the bulk radius have been 
neglected.  We then expand the graviton field $h_{\alpha\beta}$ as
\begin{equation}
h_{\alpha\beta}(x,\phi)= \sum_{n=0}^{\infty} h^{(n)}_{\alpha\beta}(x)
\frac{\chi^{n}_{G}(\phi)}{\sqrt{r_c}} \,\, ,
\end{equation}
and impose the orthonormality constraint
\begin{equation}
\int_{-\pi}^{\pi} d\phi \, e^{-2\sigma} \chi^{(m)}_{G}\chi^{(n)}_{G} = 
\delta^{mn} \,\, .
\end{equation}
The explicit forms of the KK graviton wavefunctions contain the second
order Bessel functions $J_2$ and $Y_2$, and can be found 
in~\cite{Davoudiasl:1999jd,Davoudiasl:2000wi}.  Expressing the graviton
masses as $m^{G}_{n}=x^{G}_{n} k e^{-kr_c \pi}$, we find the numerical
values $x^{G}_{1} \simeq 3.83$, $x^{G}_{2} \simeq 7.02$, $x^{G}_{3} 
\simeq 10.17$ 
for the first few states which are given by $J_1(x_n^G)=0$;
notice that $m^{G}_{n} > m^{A}_{n}$.  The couplings 
of the KK gravitons to fermions are given by
\begin{equation}
C^{mnq}_{f\bar{f}G} = \int_{-\pi}^{\pi} d\phi \frac{e^{\sigma}f^{(m)}
f^{(n)} \chi^{(q)}_{G}}{\sqrt{kr_c}} \,\, .
\end{equation}
The $C^{00n}_{f\bar{f}G}$ are presented in Fig.~\ref{FGcoups} as
functions of $\nu$.  These couplings are relatively small, particularly when 
$\nu \leq -0.5$; this, and their large mass, render the KK gravitons 
unimportant in our analysis.

%\vspace*{-2.0cm}
\noindent

We now have the tools necessary to build the SM within the RS framework.  
In the next section we will discuss the diagonalization of the $t$ and $b$ 
quark mass matrices, and the KK contributions to the $\rho$ parameter.  
To whet the reader's appetite, we note that the off-diagonal mass matrix 
elements $b_{mn}$, with $m,n \neq 1$ and $m \neq n$, range from $\approx 10$
when $\nu=-0.4$ to $\approx 700$ when $\nu=-0.6$.  This induces large mixing
between the zero mode top quark and its KK tower, and creates couplings
between the zero mode $b$ quark and the top quark KK states.  The large mass 
splitting between these states results in drastic alterations of $\rho$, and 
renders the placement of third generation quarks in the bulk
inconsistent with EW measurements for a wide range of KK masses.

\section{Fermion Mixing and the $\rho$ Parameter}

Since the top quark is quite heavy, with a mass $m_t \approx 175$ GeV, 
we expect the mixing between it and its KK tower to be stronger than that
for the lighter fermions, and we focus
here upon it and its isodoublet partner, the bottom quark.  As
mentioned previously, after performing the KK reduction of the 5-dimensional
fermion field we obtain a chiral zero mode and a vector-like KK tower; this
spectrum is presented pictorially for the top quark in Table~\ref{spectable}.
\vspace{0.5cm}
\begin{table}[h]
\centering
\begin{tabular}{|c|c||c|c|} \hline \hline
\multicolumn{2}{|c||}{Doublet} &
\multicolumn{2}{c|}{Singlet} \\ \hline
\vdots & \vdots & \vdots & \vdots \\
$T^{(2)}_{L}$ & $T^{(2)}_{R}$ & $t^{(2)}_{L}$ & $t^{(2)}_{R}$ \\
$T^{(1)}_{L}$ & $T^{(1)}_{R}$ & $t^{(1)}_{L}$ & $t^{(1)}_{R}$ \\
$T^{(0)}_{L}$ & $X$ & $X$ & $t^{(0)}_{R}$ \\ \hline
\end{tabular}
\caption{Abbreviated list of the top quark KK states.  The subscripts $L$ and
$R$ denote left-handed and right-handed fields.  $SU(2)_{L}$ doublets are 
denoted by capital $T$ and singlets by lower case $t$.  An $X$ at a location
in the table indicates that the state does not exist due to the orbifold
symmetry.}
\label{spectable}
\end{table}

\noindent
We have exchanged the left-handed $SU(2)_L$ Weyl singlets introduced in the
preceding section for their right-handed conjugates, but the reader should 
remember that these states are still described
by the 5-dimensional wavefunctions $f^{(n)}_{L} (\phi)$.  The top quark mass
matrix receives two distinct contributions: the diagonal KK couplings between 
the doublet tower states and the singlet tower states, and 
off-diagonal mixings between
the left-handed doublets and right-handed singlets arising from the Higgs
coupling on the TeV-brane.  We present below the mass matrix with only
the zero modes and the first two KK levels included; the infinite-dimensional
case can be obtained by a simple generalization.  Working in the weak 
eigenstate basis defined
by the vectors $\Psi^{t}_L = \left( T^{(0)}_{L},T^{(1)}_{L},t^{(1)}_{L},
T^{(2)}_{L},t^{(2)}_{L} \right)$
and $\Psi^{t}_R = \left( t^{(0)}_{R},t^{(1)}_{R},T^{(1)}_{R},t^{(2)}_{R},
T^{(2)}_{R} \right)$, we find
\begin{equation}
{\cal M}_{t} = \left( \begin{array}{ccccc}
   m_{t,0} & \sqrt{f}m_{t,0} & 0 & -\sqrt{f}m_{t,0} & 0 \\
   \sqrt{f}m_{t,0} & fm_{t,0} & m_1 & -\sqrt{f}m_{t,0} & 0 \\
   0 & m_1 & 0 & 0 & 0 \\
   -\sqrt{f}m_{t,0} & -fm_{t,0} & 0 & fm_{t,0} & m_2 \\
   0 & 0 & 0 & m_2 & 0 \end{array} \right) \,\, .
\end{equation}
$m_{t,0}$ is the mass of the zero mode in the infinite KK mass limit, and
$m_1$ and $m_2$ are the masses of the first two KK fermion states in the 
limit of vanishing Higgs couplings; $f$ is the mixing strength introduced in
the previous section.  We fix $m_{t,0}$ by demanding that the
lowest lying eigenvalue of this matrix reproduce the measured top quark mass, 
$m_t = 174.3$ GeV.  Due to the large values of the off-diagonal elements,
we diagonalize this mass matrix numerically, rather than
analytically to ${\cal O}(m_{t,0}/m_1)$.  We must necessarily truncate the
KK expansion at some level; we have performed our analysis twice, once
including only the first KK level and once keeping the first two levels, and
have checked that adding more states only strengthens our conclusions.  
We examine the parameter region $-0.3 \gsim \nu \gsim -0.55$; the range 
$\nu > -0.3$ is strongly constrained by contact interaction searches at 
LEP~\cite{Davoudiasl:2000wi}, and the values $\nu \lsim -0.55$ are prohibited 
by extrapolation of the results obtained below. This is essentially the same 
region studied in~\cite{delAguila:2000kb,delAguila:2000fg}, where it was shown 
that the LHC will be able to probe the shift in the $Zt\bar{t}$ coupling for 
fermion KK mass values $m_1 \lsim 15$ TeV.  We will find that the region where
$\nu \leq -0.3$ and $m_1 \lsim 30-100$ TeV is already disfavoured by current 
measurements.

The Lagrangian containing the top quark mass terms and its interactions with
the $Z$ and $W^{\pm}$ gauge bosons is
\begin{eqnarray}
{\cal L}&=&\left( \bar{\Psi}^{t}_L {\cal M}_{t} \Psi^{t}_{R} +{\rm h.c.} 
\right) +
\bar{\Psi}^{t}_{L} \not\!{Z} C^{Z}_{t,L} \Psi^{t}_{L} + \bar{\Psi}^{t}_{R} 
\not\!{Z} C^{Z}_{t,R} \Psi^{t}_{R} \nonumber \\ & &
+ \bar{\Psi}^{t}_{L} \not\!{W}^{-} 
C^{W}_{L} \Psi^{b}_{L} + \bar{\Psi}^{t}_{R} \not\!{W}^{-} C^{W}_{R} 
\Psi^{b}_{R} + \,... + {\rm h.c.}\,\, .
\end{eqnarray}
We have introduced the basis $\Psi^{b}_{L}$ and $\Psi^{b}_{R}$ for the bottom
quark in analogy with those for the top quark.
The $C^{i}_{j}$ are matrices containing the couplings of the various top quark
states to the $Z$ and $W^{\pm}$; letting $g$ denote the SM electroweak 
coupling, $c_{W}$ the cosine of the weak mixing angle, and $g_L$ and $g_R$
the couplings of the usual left-handed and right-handed SM fermions to the 
$Z$ boson, we find
\begin{eqnarray}
C^{Z}_{t,L} &=& \frac{g}{c_{W}} \, {\rm diag} \left(g_L,g_L,g_R,g_L,g_R 
\right) \,\, , \nonumber \\ 
C^{Z}_{t,R} &=& \frac{g}{c_{W}} \, {\rm diag} \left(g_R,g_R,g_L,g_R,g_L 
\right) \,\, , \nonumber \\ 
C^{W}_{L} &=& \frac{g}{\sqrt{2}} \, {\rm diag} \left(1,1,0,1,0 \right) \,\, ,
\nonumber \\ 
C^{W}_{R} &=& \frac{g}{\sqrt{2}} \, {\rm diag} \left(0,0,1,0,1 \right) \,\, .
\end{eqnarray}
In obtaining these matrices we have treated the $T^{(n)}_{R}$ as $SU(2)_L$
doublets and the $t^{(n)}_{L}$ as singlets as denoted in Table \ref{spectable}.
We diagonalize ${\cal M}_{t}$
with the two unitary matrices $U^{t}_L$ and $U^{t}_R$,
\begin{equation}
{\cal M}^{D}_{t} = U^{t}_L {\cal M}_{t} \left(U^{t}_{R}\right)^{\dagger} \,\, .
\end{equation}
Diagonalization of the matrix ${\cal M}_{t} {\cal M}^{\dagger}_{t}$ determines
$U^{t}_L$ up to an overall phase matrix, while diagonalization of 
${\cal M}^{\dagger}_{t} {\cal M}_{t}$ similarly fixes
$U^{t}_R$.  The mass eigenstate basis is obtained by multiplication of
the weak eigenstate basis by the appropriate transformation matrix:
\begin{equation}
\Psi^{t}_{L} \rightarrow U^{t}_{L} \Psi^{t}_{L} \,\,\,\, ,
\Psi^{t}_{R} \rightarrow U^{t}_{R} \Psi^{t}_{R} \,\, .
\end{equation}
The coupling matrices undergo a similar shift,
\begin{eqnarray}
C^{Z}_{t,L} &\rightarrow& U^{t}_{L} C^{Z}_{t,L} \left(U^{t}_{L} 
\right)^{\dagger} \,\, , \nonumber \\
C^{Z}_{t,R} &\rightarrow& U^{t}_{R} C^{Z}_{t,R} \left(U^{t}_{R} 
\right)^{\dagger} \,\, , \nonumber \\
C^{W}_{L} &\rightarrow& U^{t}_{L} C^{W}_{L} \left(U^{b}_{L} 
\right)^{\dagger} \,\, , \nonumber \\
C^{W}_{R} &\rightarrow& U^{t}_{R} C^{W}_{R} \left(U^{b}_{R} 
\right)^{\dagger} \,\, .
\end{eqnarray}
We have also implicitly performed an identical diagonalization of the
bottom quark mass matrix.

\begin{figure}[htbp]
\centerline{
\psfig{figure=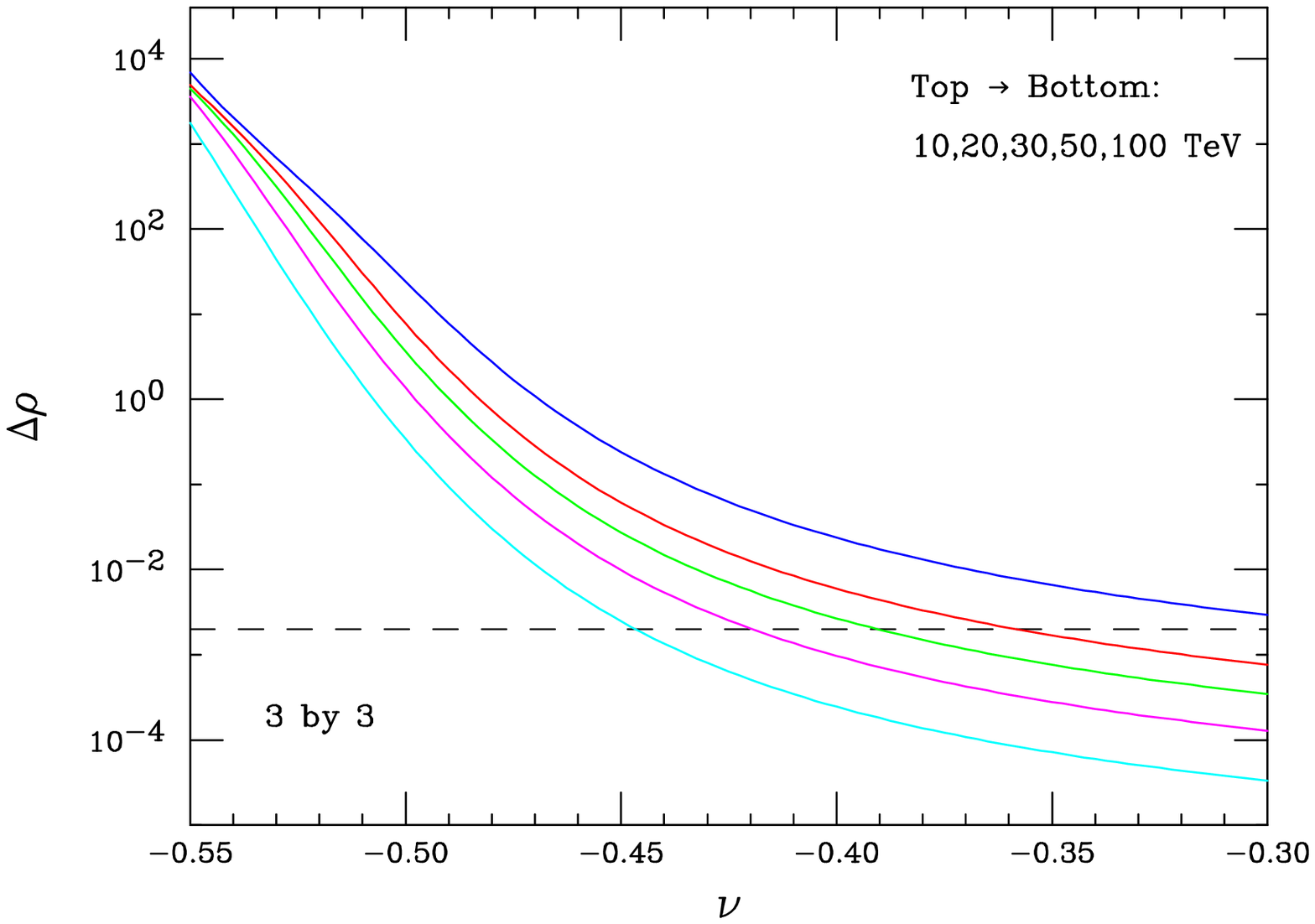,height=8.5cm,width=10.5cm,angle=0}}
\vspace{0.5cm}
\centerline{
\psfig{figure=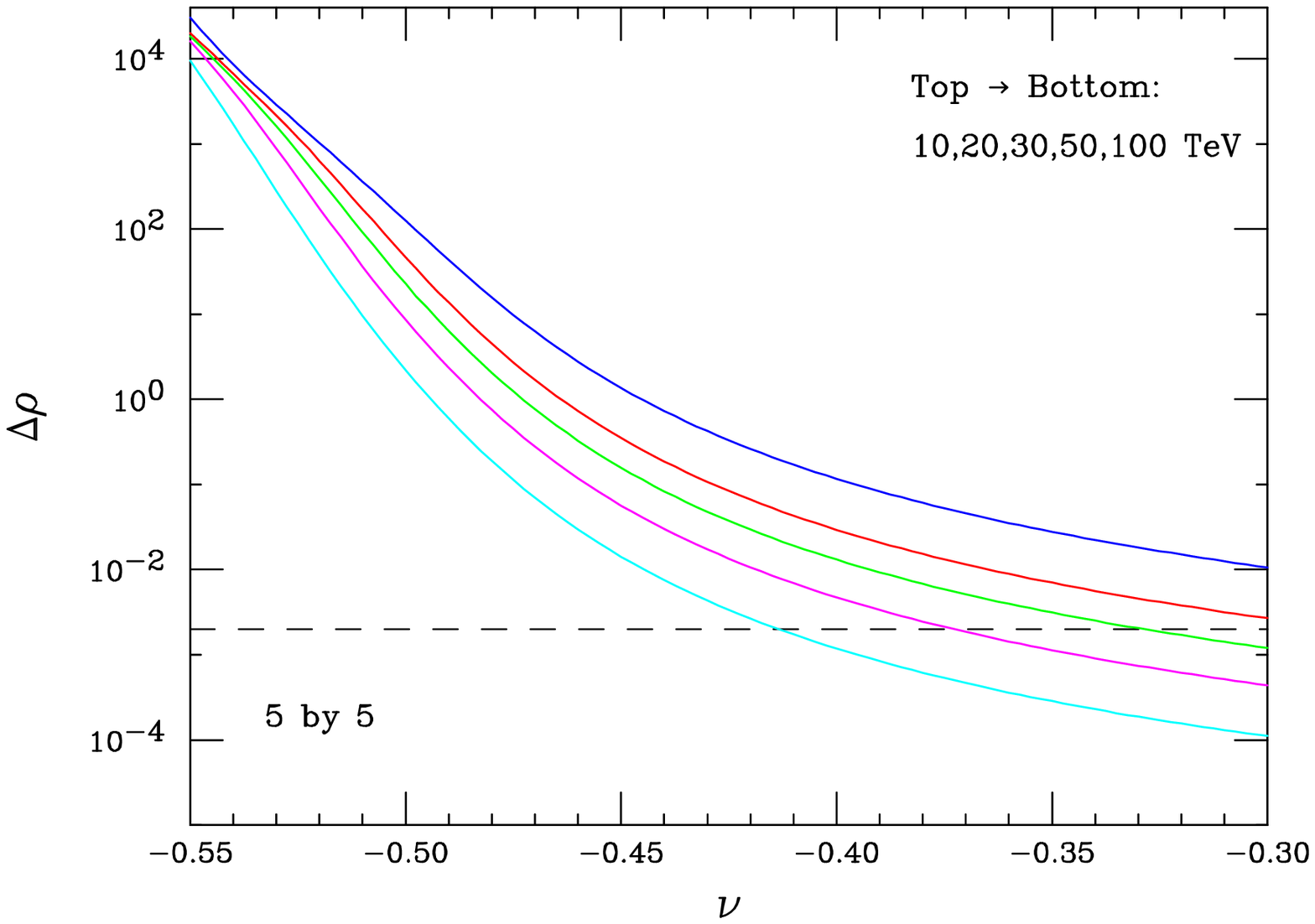,height=8.5cm,width=10.5cm,angle=0}}
\caption{Contributions to $\Delta \rho$ from the zero modes and first KK level 
(top), and from the zero modes and first two KK levels (bottom).  The dashed
black line indicates where $\Delta \rho = 2 \times 10^{-3}$.  The various curves
correspond to when the masses of the first fermion KK excitation is taken to be
$m_1=10, 20, 30, 50, 100$ TeV, from top to bottom.}
\label{rho}
\end{figure}

This procedure induces off-diagonal elements in both the $Z$ and $W^{\pm}$
coupling matrices; consequently, fermions of widely varying masses enter the
vacuum polarization graphs contributing to the $Z$ and $W^{\pm}$ self 
energies.  Such a scenario typically generates unacceptable contributions to
the $\rho$ parameter~\cite{Veltman:1977kh}, defined as
\begin{equation}
\rho = \frac{\Pi_{W}\left(q^2 =0 \right)}{M_{W}^2} - 
\frac{\Pi_{Z}\left(q^2 =0 \right)}{M_{Z}^2} \,\, ,
\end{equation}
where $\Pi_{X} \left(q^2 \right)$ is the $X$ boson self energy function.  We
set $\Delta \rho=\rho -\rho_{SM}$, where $\rho_{SM}$ is the contribution
from the SM $(t,b)$ doublet, and calculate $\rho$ for our two cases: 
once including the shifted top and bottom quark zero modes and the first KK 
level only, and once
including the zero modes and the first two KK states.  $\Delta \rho$ is then
a measure of the deviation from the SM prediction.  The results are presented 
in Fig.~\ref{rho} as 
functions of $\nu$ for several choices of $m_1$.  The 95\% CL exclusion
limit \cite{Langacker:2001ij} of $\Delta\rho\lsim 2\times 10^{-3}$ is also
indicated.  Notice that $\Delta \rho$ increases when we add the second 
KK level in our analysis; thus adding more states only
increases $\Delta \rho$ further, and our neglect of these higher modes is 
justified.  It is
clear from the lower graph in Fig.~\ref{rho} that 
consistency with the 95\% CL exclusion limit
restricts $m_1$ to the range $m_1 \gsim 25$ TeV for all values of $\nu$
in the previously allowed range,
and requires $m_1 \gsim 100$ TeV when $\nu \leq -0.4$.  When 
$\nu <-0.5$, including the range $\nu \leq -0.55$ that we have not presented,
the corrections to $\rho$ are so large that the perturbative definition of
the $Z$ and $W^{\pm}$ gauge bosons is no longer valid.  We stress that 
these restrictions are lower bounds on the actual constraints as including
more KK levels in our analysis will only strengthen these results. 
These results imply
$\Lambda_{\pi} \gsim  100$ TeV~\cite{Davoudiasl:1999tf}, with the exact 
choice depending on the value of $\nu$, to avoid unacceptable contributions to
$\Delta \rho$; the resulting hierarchy between the EW scale and the
fundamental RS scale thus strongly disfavors allowing the third generation
quarks to propagate in the bulk.

This restriction applies only to the top and bottom quarks; the first and 
second generations are much less massive, and the large mixing induced 
above by the top quark Yukawa coupling does not appear when 
considering these states.   We have numerically checked that the
contributions of bulk first and second generation quarks are consistent
with the constraints on $\Delta\rho$, and hence the 
placement of the first two generations in the bulk is 
still allowed.  But, is
such a setup motivated?  Does any interesting physics result from this
construction?  The answer to both questions is unequivocally yes, as we
will demonstrate in the next section.
%
%\vspace*{-2.0cm}
\noindent

\section{The Third Generation On the Wall and the EW Precision 
Observables}

A handful of authors have attempted to construct models explaining the quark 
and neutrino mass matrices within the framework of the RS 
model~\cite{Grossman:1999ra,big}.  These 
ideas generically require the placement of fermions at different locations in 
the 5-dimensional bulk.  We have already shown that the third generation 
quarks must lie on the TeV-brane; if we permit the first two generations to 
propagate in the bulk, can we explain the hierarchy between the Yukawa
coupling of the top quark and those of the lighter quarks?

We consider first the coupling of the Higgs to a fermion field confined to the
TeV-brane; the relevant action is
\begin{equation}
S^{wall}_{f\bar{f}H}= \lambda^{wall} \int d^{4}x \int dy \sqrt{-G} 
\left\{ H^{\dagger}
\psi_{D} \psi^{c}_{S} + {\rm h.c.} \right\} \delta \left( y-r_c \pi \right) 
\,\, ,
\end{equation}
where $\lambda^{wall}$ is the Yukawa coupling of the localized fermion, 
chosen to be of 
${\cal O}(1)$.  To derive the 4-dimensional action we must rescale 
$\psi \rightarrow e^{3\sigma /2}\psi$ and $H \rightarrow e^{\sigma} H$ as
before.  The mass of this field is then $m^{wall} = \lambda^{wall} v/\sqrt 2$.
We have shown in Eq.~\ref{yukawa} that the 4-d Yukawa coupling that 
determines the zero-mode mass of a bulk fermion is
\begin{equation}
\lambda^{bulk} = \frac{\lambda^{'} \left(f^{(0)}_{L}(\pi) \right)^2 
e^{kr_c \pi}}{kr_c} \,\, . 
\end{equation}
We now assume that the \emph{fundamental} coupling that enters the 5-d 
bulk fermion action, $\lambda^{'}$, is also of order unity as is 
$\lambda^{wall}$.  The factor 
$ \left(f^{(0)}_{L} \right)^2 e^{kr_c \pi} /kr_c$ then suppresses 
$\lambda^{bulk}$ with respect to $\lambda^{wall}$; we find
$\lambda^{bulk} \approx \left( 10^{-1}-10^{-2} \right) \lambda^{wall}$ when
$-0.55 \lsim \nu \lsim -0.35$, using the zero mode wavefunction given in
Eq.~\ref{fermzero}.  Choosing a value of $\nu$ in this region, 
ameliorates the
hierarchy between the second and third generation quark Yukawa couplings.  To 
be more explicit, with the top quark on the TeV brane and $\lambda^{wall}$ 
of order unity, we expect a top mass near its experimental value. On the 
other hand, for the charm quark in the bulk, we expect a much smaller mass 
even if the bulk Yukawa coupling $\lambda'$ is also of order unity. As an 
example, assuming $\lambda'=\lambda^{wall}$ and taking $\nu=-0.5$ one obtains 
$m_t/m_c \simeq 2\pi kr_c \simeq 70$ which is within a factor of 2 to 3 of the 
experimental value. A similar argument applies to the $m_b/m_s$ ratio. The 
localization of the third generation on the TeV brane while keeping the 
first two 
in the bulk may thus help explain the fermion mass hierarchy.  We do 
not attempt here to build a more detailed flavor model incorporating
the off-diagonal CKM matrix elements, but instead examine the consequences
of this simple situation.  We focus on the region $-0.6 \leq \nu \leq -0.3$, 
extending slightly for completeness the range preferred by the quark Yukawa
hierarchy.  We study next the effects of KK gauge boson mixing 
on the EW precision
observables; we will find that large mixing similar to that appearing in the
top quark mass matrix relaxes the upper bound on the Higgs boson mass
obtained in the standard EW fit \cite{peskinwells}.

The mass terms for the $W^{\pm}$ and $Z$ can be obtained using 
Eq.~\ref{genmass} as a template; we find
\begin{equation}
S_{mass} = \sum_{m,n=0}^{\infty} a_{mn} \int d^{4}x \left( m_{W,0}^2 
W^{+(m)}_{0} W^{-(n)}_{0} +\frac{1}{2}m_{Z,0}^2 Z^{(m)}_{0} Z^{(n)}_{0} 
\right) \,\, ,
\end{equation}
where the $a_{mn}$ are given by Eq.~\ref{mixterms} and we have for notational
simplicity omitted the Lorentz indices of the gauge fields.  The resulting 
$W^{\pm}$ mass matrix is 
\begin{equation}
{\cal M}_{W}^2 = m_{1}^2 \left( \begin{array}{cccc}
               a_{11}x_{W} & a_{12}x_{W} & a_{13}x_{W} & \ldots \\
               a_{12}x_{W} & b_{1}^2 +a_{22}x_{W} & a_{23}x_{W} & \ldots \\
               a_{13}x_{W} & a_{23}x_{W} & b_{2}^2 +a_{33}x_{W} & \ldots \\
               \vdots & \vdots & \vdots & 
               \end{array} \right) \,\, ,
\end{equation}
where $m_{1}$ is the first KK gauge mass, $x_{W}=m_{W,0}^2 /m_{1}^2$,
and $b_{i} =m_{i} /m_{1}$ is the ratio of the $i$th KK mass to
the first.  The mass matrix for the $Z$ is obtained by substituting
$m_{W,0} \rightarrow m_{Z,0}$.  The subscripts on the fields $W^{\pm}_{0}$ and 
$Z_{0}$, and on the masses $m_{W,0}$ and $m_{Z,0}$, indicate that these
are not the physical fields and masses; they are the zero-mode fields
and masses in the infinite KK limit.  To obtain the physical spectrum
we must diagonalize the mass matrices while respecting the
appropriate constraints; these are the same as those
developed for the interpretation of precision measurements at the 
$Z$-pole~\cite{Altarelli:hv}.  This prescription states that the following
quantities are inputs to radiative corrections and to fits to the precision
EW data: $\alpha$ as measured in Thomson scattering, $G_F$ as defined
by the muon lifetime, $M_Z$ as determined from the $Z$ line shape, $m_t$ as 
measured at the Tevatron, and $m_H$, which is currently a free parameter.  
All other observables, such as the $W^{\pm}$ mass, $M_W$, and the width for 
the decay $Z \rightarrow l^+ l^-$, $\Gamma_l$, are derived from these 
measured quantities; we must compare the RS model predictions for these 
parameters with the values obtained by experiment.

We examine the six relatively uncorrelated observables $M_W$, 
${\rm sin}^{2}\theta_{\rm eff}$, $\Gamma_l$, $R_b$, $R_c$, and 
${\rm sin}^{2}\theta_{\rm \nu N}$, and discuss in detail our procedure for 
deriving the RS model predictions for these quantities and then compare these 
predictions to the measured values.  We consider tree level KK and loop level 
SM contributions to these observables, and assume that contributions from KK 
loops are higher order and therefore negligible.  Our analysis differs
slightly from those performed in models with KK gauge bosons arising from
${\rm TeV}^{-1}$-sized extra dimensions~\cite{Masip:1999mk,
Rizzo:1999br}.  Here, the parameters $a_{mn}$ of Eq.~\ref{mixterms} 
which enter the mass matrices are rather large;
the $a_{1n}$, with $n>1$, have the approximate value $\sqrt{2\pi kr_c} \approx
8.4$, while the elements $a_{mn}$, with $m,n >1$ and $m \neq n$, have the
approximate value $2\pi kr_c \approx 71$.  Although the ratios
$x_{W(Z)}=m_{W,0(Z,0)}^2 /m_{1}^2$ that appear in the mass matrices 
may be small, 
they are multiplied by these large coefficients, and to avoid errors we 
diagonalize the matrices and handle shifts of the precision observables 
numerically to all orders in $x_{W,Z}$, rather than performing the
analysis analytically to 
${\cal O}(x_{W,Z})$.  This necessitates a truncation of the mass matrices; we 
work with $30 \times 30$ matrices, and have verified that increasing the size 
to $60 \times 60$ produces a negligible change in our results.

We first determine the parameter $m_{Z,0}$ by diagonalizing the $Z$ mass 
matrix and demanding that the lowest eigenvalue reproduce the measured
$Z$ mass, $M_Z$ .  Armed with $m_{Z,0}$, we consider next the muon
lifetime, through which the input parameter $G_F$ is defined.  The
relevant decay is $\mu^- \rightarrow e^- \nu_e \bar{\nu}_{\mu}$.  In the SM
this proceeds at tree level through $W$ exchange; it proceeds here through
the exchange of the entire $W^{(n)}$ KK tower.  $G_F$ therefore becomes
\begin{equation}
\frac{G_F}{\sqrt{2}} = \frac{g^2}{8M_{W}^2} + \frac{g_{0}^2}{8}\sum_{n} 
\frac{c_n}{m_{n}^2} \,\, ,
\label{GFdef}
\end{equation}
where the first term arises from the exchange of the zero mode and the second
term from the higher KK states, and the $c_n$ encapsulate the couplings of
the KK gauge states to zero mode leptons.  Some clarification of this 
expression is required.  $g$ is the coupling of the \emph{physical} $W$ 
obtained \emph{after} diagonalization, whereas $g_0$ is the coupling that 
appears in the Lagrangian \emph{before} diagonalization.  To make this 
distinction explicit we express $g$ as
\begin{equation}
g=g_{0} \left\{ 1-G(m_{W,0}) \right\} \,\, ,
\end{equation}
where $G(m_{W,0})$ accounts for the admixture of KK states in the physical
$W^{(0)}$ boson.  
After EW symmetry breaking, $g_{0}^2 = 4\pi\alpha/s_{w,0}^2$, 
where $s_{w,0}$ is the sine of the weak mixing angle obtained before
including mixing effects: $s_{w,0}^2 = 1-m_{W,0}^2 /m_{Z,0}^2$.  Substituting 
these relations into Eq.~\ref{GFdef}, we arrive at the condition
\begin{equation}
1 = \frac{\pi\alpha}{\sqrt{2}G_F \,  M_{W}^2 s_{w,0}^2} \left\{1-G(m_{W,0}) 
\right\}^2 +\frac{\pi\alpha}{\sqrt{2}s_{w,0}^2} H(m_{W,0}) \,\, ,
\label{GFdefnew}
\end{equation}
where we have introduced the dimensionless quantity
\begin{equation}
H(m_{W,0})=\frac{1}{G_F}\sum_{n} \frac{c_n}{m_{n}^2} \,\, .
\end{equation}
In the SM, after radiative corrections are included,
\begin{equation}
\frac{\pi\alpha}{\sqrt{2}G_F} \rightarrow \frac{\pi\alpha}{\sqrt{2}G_F \left(
1-\Delta r \right)} = m_{W,SM}^2 \left[1-\frac{m_{W,SM}^2}{M_{Z}^2} \right]
\,\, .
\end{equation}
To incorporate radiative corrections in our analysis, we make this
substitution in Eq.~\ref{GFdefnew}, and use the $m_{W,SM}$ calculated by
ZFITTER~\cite{Bardin:1999yd} using $M_Z$ as an input.  We find the relation
\begin{equation}
1=\frac{m_{W,SM}^2 \left[1-\frac{m_{W,SM}^2}{M_{Z}^2} \right]}{M_{W}^2 \left[
1-\frac{M_{W}^2}{M_{Z}^2} \right]} \left\{ 1-G(m_{W,0})\right\}^2 
+\frac{\pi\alpha}{\sqrt{2}\left(1-\frac{m_{W,0}^2}{m_{Z,0}^2} \right)} 
H(m_{W,0}) \,\, .
\end{equation}
The only unknown quantity in this equation is $m_{W,0}$; the physical 
$W^{\pm}$ mass, $M_W$, is derived from $m_{W,0}$ through diagonalization of 
the $W^{\pm}$ mass
matrix.  We now scan over $m_{W,0}$ until we find a solution to this 
equation; the $m_{W,0}$ that furnishes this solution also predicts a $M_W$ 
that can be compared with experiment.

We next compute the KK contributions to the effective coupling 
${\rm sin}^{2}\theta_{eff}$, which appears in the dressed $Zl^+ l^-$ 
vertex~\cite{Bardin:1999yd}.
In the SM,
\begin{equation}
{\rm sin}^{2}\theta_{eff,SM} =\kappa^Z {\rm sin}^{2}\theta_{w,SM} \,\, ,
\end{equation}
where $\theta_w$ is the on-shell weak mixing angle, ${\rm sin}^{2}\theta_{w,SM}
= 1-m_{W,SM}^2 /M_{Z}^2$, and $\kappa^Z$ contains a subset of the radiative 
corrections 
to the decay $Z \rightarrow l^+ l^-$.  In the RS model, the weak mixing angle 
that appears in the $Zl^+ l^-$ vertex is $s_{w,0}$; this is unaffected by 
diagonalization because $s_{w,0}$ enters the coupling of every KK 
excitationstate.  The 
RS model expression for ${\rm sin}^{2}\theta_{eff}$ is
\begin{equation}
{\rm sin}^{2}\theta_{eff} =\kappa^Z s_{w,0}^2 = {\rm sin}^{2}\theta_{eff,SM} 
\left( 
\frac{1-\frac{m_{W,0}^2}{m_{Z,0}^2}}{1-\frac{m_{W,SM}^2}{M_{Z}^2}} \right)
\,\, ,
\end{equation}
where in the last step we have incorporated the ZFITTER predictions for 
${\rm sin}^{2}\theta_{eff,SM}$ and $m_{W,SM}$ to correctly account for the SM 
radiative corrections. 

The shifts of the remaining observables occur in a similar fashion as in 
the two examples given above, and hence
we discuss them only briefly here.  The width of the decay
$Z \rightarrow \bar{f}f$ is
\begin{equation}
\Gamma_f = \frac{g^2 M_Z}{96\pi c_{w}^2} C_f \left\{ \left[ 1-4|Q_f| 
{\rm sin}^{2}\theta_{eff} +8 Q_{f}^2 {\rm sin}^{4}\theta_{eff} \right] 
\left(1+\frac{2 m_{f}^2}{M_{Z}^2} \right) -3 \frac{m_{f}^2}{M_{Z}^2} \right\}
\,\, ,
\end{equation}
where $C_f$ encapsulates kinematic factors, color sums for final state quarks,
and factorizable radiative corrections~\cite{Bardin:1999yd}.  This formula is 
valid in both the SM and the RS model, with the proviso that in the RS case
$g$ describes the coupling of the $Z^{(0)}$ obtained after 
diagonalization, $c_{w} \rightarrow c_{w,0}$ and  
${\rm sin}^{2}\theta_{eff}$ is given by that described in the 
previous paragraph.  Our
previous results can be adapted to compute the shift in $\Gamma_l$.  The 
change in the ratio of the $Z \rightarrow
\bar{c}c$ width to the total hadronic width, $R_c = \Gamma_c / \Gamma_h$, can 
also be computed by following the outline presented for calculating the 
$\Gamma_l$ shift.  The derivation of the shift in $R_b = \Gamma_b / \Gamma_h$ 
proceeds similarly, except that the couplings of the higher gauge KK modes 
to the brane localized bottom quarks are those presented in 
Eqs.~\ref{wallcoup1} and~\ref{wallcoup2}.  Finally, 
${\rm sin}^{2}\theta_{\rm \nu N}$ is determined experimentally through the
measurement of $R$, which is the following ratio 
of neutrino-nucleon neutral and charged current scattering events:
\begin{equation}
R = \frac{\sigma^{\nu}_{NC}-\sigma^{\bar{\nu}}_{NC}}{\sigma^{\nu}_{CC}-
\sigma^{\bar{\nu}}_{CC}} \,\, .
\end{equation}
It becomes
\begin{equation}
R = \frac{1}{2} -{\rm sin}^{2}\theta_{\nu N} 
\end{equation}
at tree level in the SM, where the $W^{\pm}$ and $Z$ coupling constants have 
cancelled in the ratio.  When RS corrections are included,  
the gauge boson couplings no longer cancel because of different mixing effects 
in the $W^{\pm}$ and $Z$ mass matrices, and ${\rm sin}^{2}\theta_{\nu N} 
\rightarrow s_{w,0}^2$.  Again, these corrections to
${\rm sin}^{2}\theta_{\rm \nu N}$ can be easily
obtained from our above results.
 
\begin{figure}
\centerline{
\hspace*{-0.8cm}
\psfig{figure=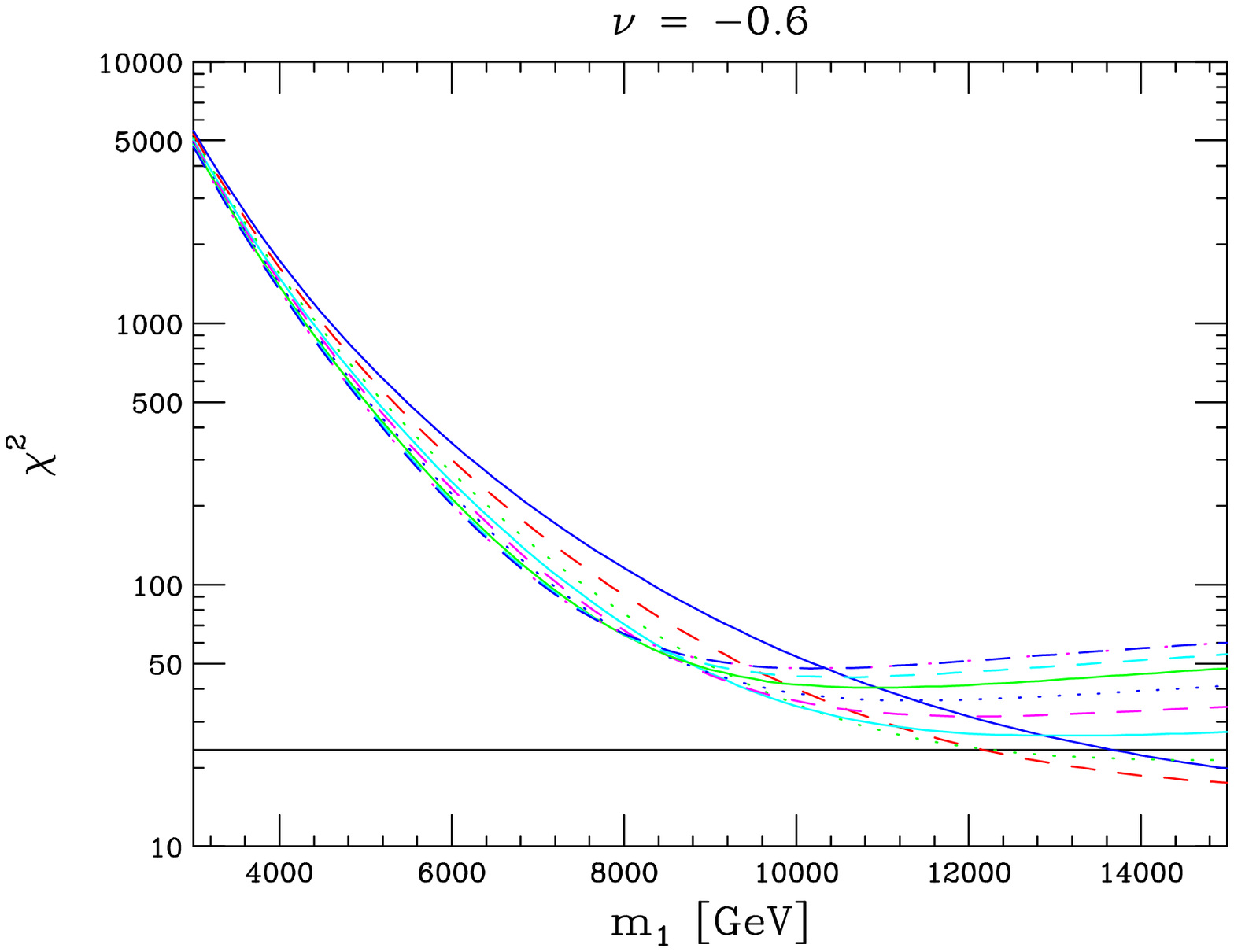,height=7.2cm,width=8.5cm,angle=0}
\hspace*{0.2cm}
\psfig{figure=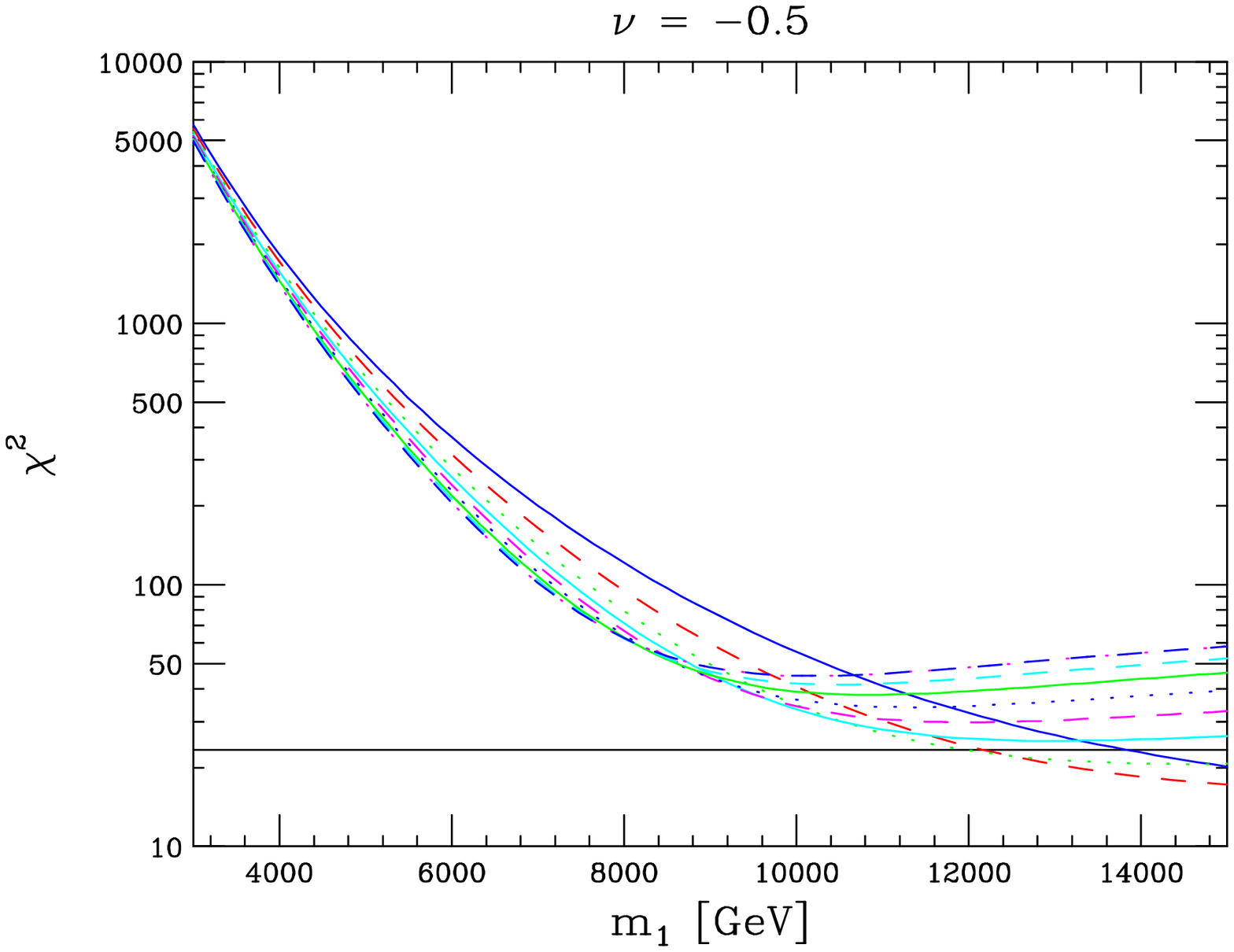,height=7.2cm,width=8.5cm,angle=0}}
\vspace*{0.5cm}
\centerline{
\hspace*{-0.8cm}
\psfig{figure=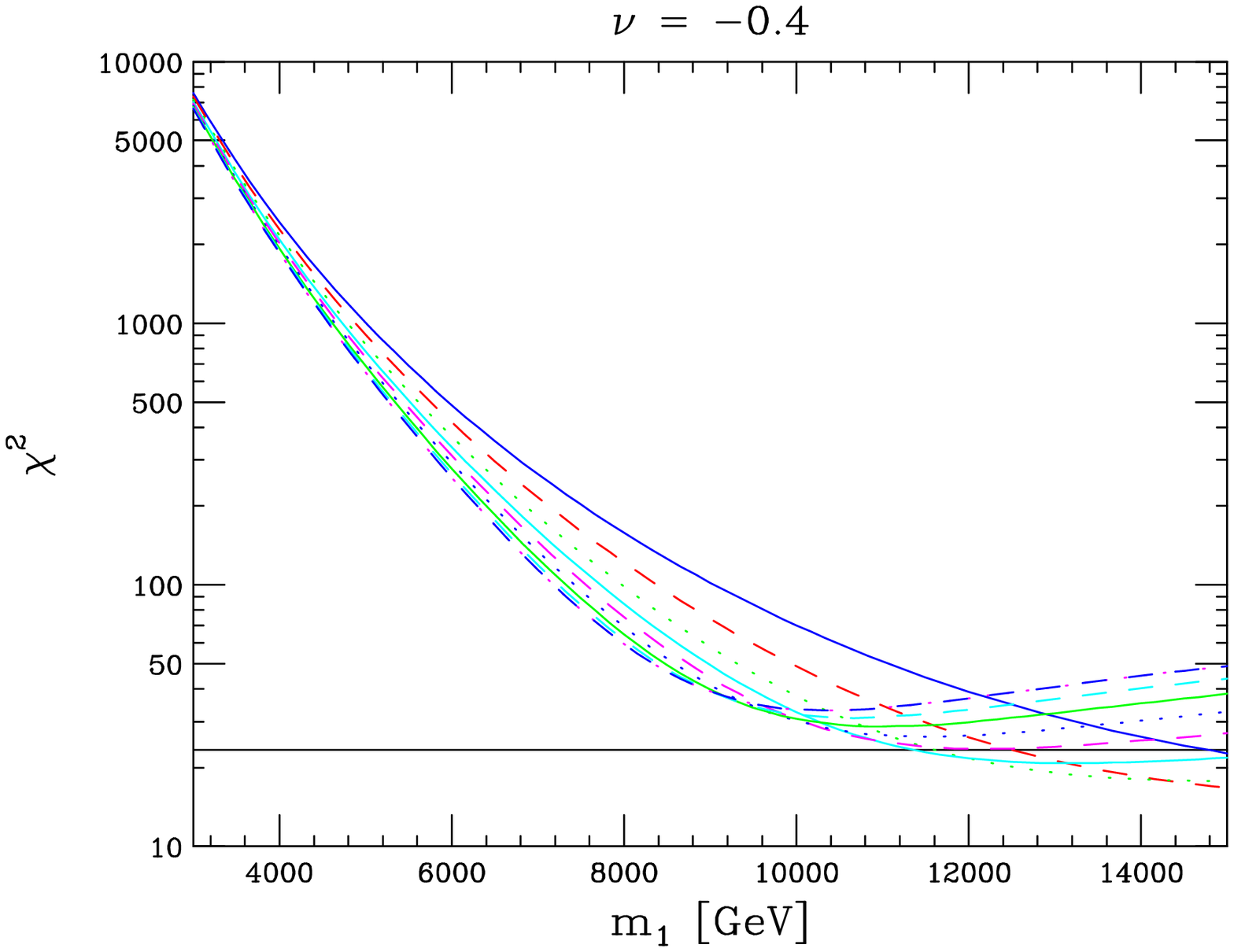,height=7.2cm,width=8.5cm,angle=0}
\hspace*{0.2cm}
\psfig{figure=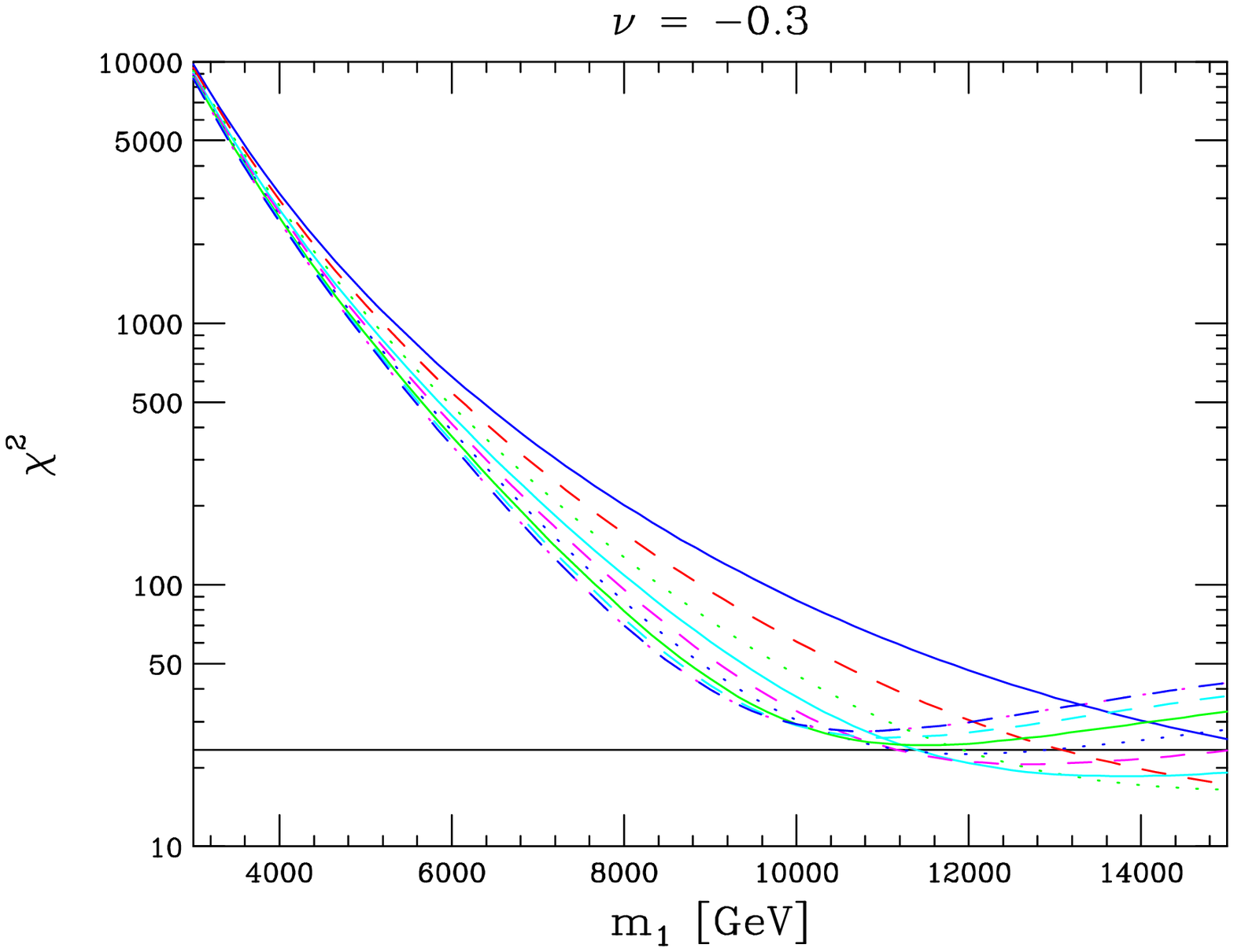,height=7.2cm,width=8.5cm,angle=0}}
\vspace*{0.5cm}
\caption{$\chi^2$ values obtained in the fit to the EW precision
data as a function of $m_1$ for four choices of $\nu$,
the fermion bulk mass parameter.  The solid black line indicates where
$\chi^2 = 23.3$, the value at which the 95\% CL is reached.  The colored
curves are the RS model fit results for different Higgs boson masses; from
top to bottom, on the left of each plot, the lines indicate $m_H = 115$, 
$200$, $300$, \ldots, $1000$ GeV. }
\label{EWchi}
\end{figure}

\begin{figure}[htbp]
\centerline{
\psfig{figure=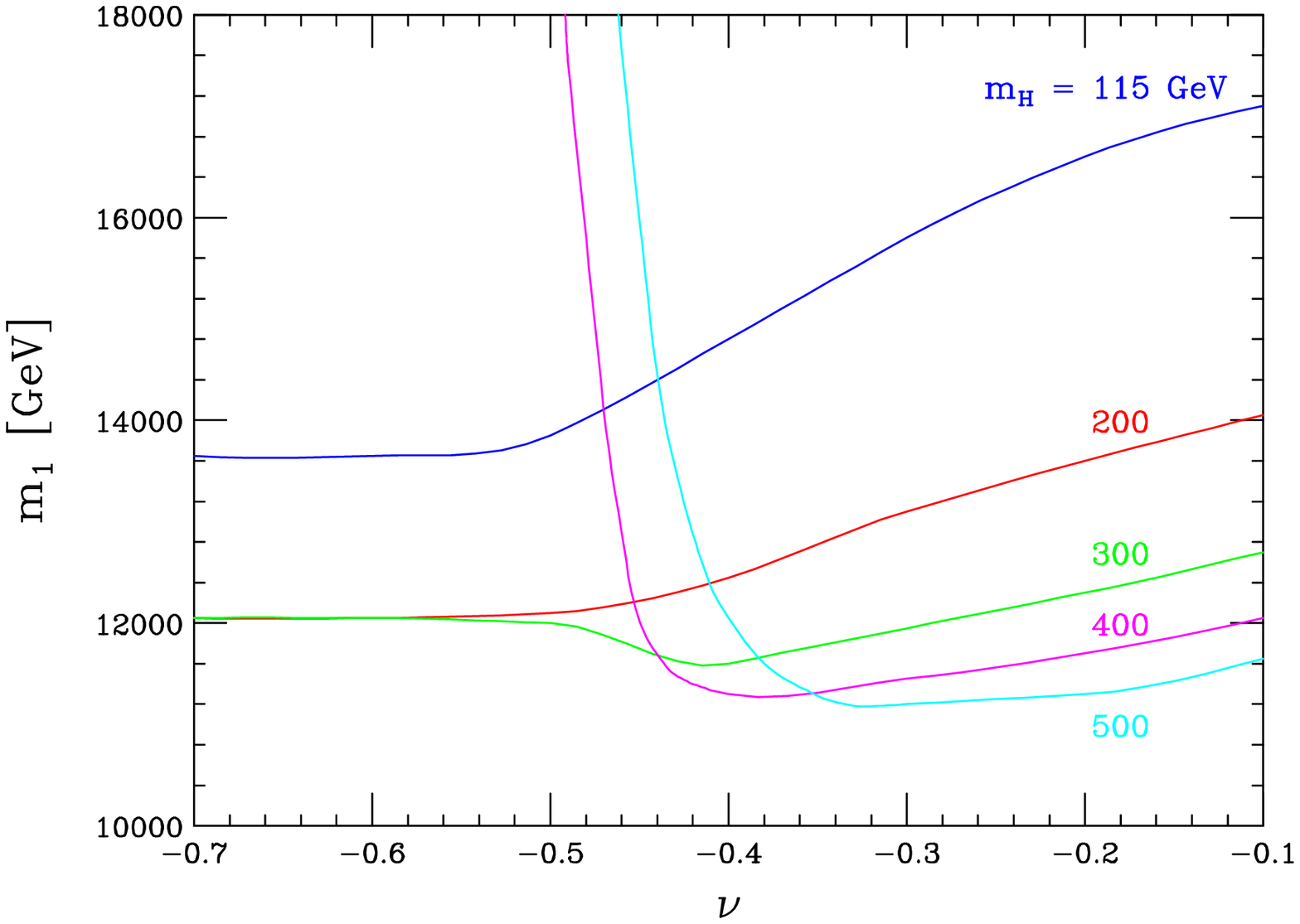,height=7.7cm,width=9.5cm,angle=0}}
\caption{Lowest value of $m_1$ that fits the EW data as a function of
$\nu$, for five representative choices of $m_H$.}
\label{lowm1}
\end{figure}

Having computed these corrections, we can now compare the RS model predictions 
for the precision observables with the values actually measured.  We perform a 
$\chi^2$ fit to the data, with the Higgs boson mass $m_H$ and the first KK 
gauge mass $m_1$ as free parameters.  The LEP Electroweak Working Group has 
quoted an upper limit on the Higgs mass in the SM of $m_H < 222$ GeV at the 
95\% confidence level~\cite{Abbaneo:2001ix}, which we find corresponds to 
$\chi^2 = 23.3$.  Following~\cite{Rizzo:1999br}, we normalize our results
by choosing this $\chi^2$ value as our benchmark; we claim that the 
predictions are disfavoured at the 95\% CL if $\chi^2 > 23.3$, and that the 
model fits the precision data otherwise.  We use the input parameter values
\begin{eqnarray}
M_Z &=& 91.1875 \,\, {\rm GeV} \,\, , \nonumber \\
G_F &=& 1.16637 \times 10^{-5} \,\, {\rm GeV}^{-2} \,\, , \nonumber \\
\alpha(m_e) &=& 1/137.036 \,\, ,
\end{eqnarray}
and the experimental observable values and errors
\begin{eqnarray}
M_W &=& 80.451 \pm 0.033 \,\, {\rm GeV} \,\, , \nonumber \\
{\rm sin}^{2}\theta_{eff} &=& 0.23152 \pm 0.00017 \,\, , \nonumber \\
\Gamma_l &=& 83.991 \pm 0.087 \,\, {\rm MeV} \,\, , \nonumber \\
R_b &=& 0.21646 \pm 0.00065 \,\, , \nonumber \\
R_c &=& 0.1719 \pm 0.0031 \,\, , \nonumber \\
{\rm sin}^{2}\theta_{\rm \nu N} &=& 0.2277 \pm 0.0016 \,\, ,
\end{eqnarray}
as presented in~\cite{pm,Abbaneo:2001ix}.  The results of these
fits are presented in Fig.~\ref{EWchi} as functions of both $m_H$ and
$m_1$, and for four representative values of the fermion bulk mass parameter 
$\nu$.  We have allowed $m_H$ to range from 115 GeV to 1 TeV;
higher Higgs masses are inconsistent with perturbative unitarity.  This bound
is modified slightly by KK gauge boson and graviton exchanges, but we have
neglected these effects here.  The six observables contribute to the fit
with widely varying strengths;
${\rm sin}^{2}\theta_{eff}$ is very sensitive to deviations arising from RS
physics throughout the entire $m_H$, $m_1$ region, while $R_c$ does not 
significantly affect the $\chi^2$ value for any choice of parameters.  $R_b$
is drastically altered when $\nu \leq -0.5$, where the light fermion couplings
to KK gauge states either vanish or become small, but is less affected for
larger values of $\nu$.  $M_W$, $\Gamma_l$, and ${\rm sin}^{2}\theta_{\rm 
\nu N}$ are somewhat less 
sensitive than ${\rm sin}^{2}\theta_{eff}$, and vary in relative importance
as $m_H$ and $m_1$ are changed.  The allowed values of $m_H$ vary with 
$\nu$, but it is clear from Fig.~\ref{EWchi} that for $\nu \geq -0.5$ Higgs
masses in the range $300-600$ GeV are permitted for $n=1$ KK gauge masses of
$11 \sim 15$ TeV.  A heavy Higgs has the effect of decreasing
$M_W$, while the RS mixing effects increase it, and this compensation allows 
the predicted values of $M_W$, ${\rm sin}^{2}\theta_{eff}$, and $\Gamma_l$ to 
be brought into good agreement with the measured values by tuning $m_1$.  
Shifts in $R_b$ arising from the confinement of the third generation quarks to 
the TeV-brane prevent larger values of $m_H$ from providing a good fit to the 
EW precision data.  For each choice of $\nu$ and $m_H$ there exists a range of 
allowed $m_1$ values that fits the EW precision data; the lowest allowed 
value of $m_1$ as a function of $\nu$ is presented in Fig.~\ref{lowm1} for 
several choices of $m_H$.  The drastic difference between the $m_H =300$ 
and $400$ GeV curves arises from the sharp distinction between allowed and
disallowed KK masses imposed by the cut at $\chi^2 =23.3$.  The sharp rise
for lower $\nu$ values and higher Higgs masses is due almost entirely
to $R_b$.  We present in Table~\ref{m1range} a summary of the allowed $m_1$ 
ranges for various choices of $\nu$ and $m_H$.
\vspace{0.5cm}
\begin{table}[h]
\centering
\begin{tabular}{|r|r|r|r|r|} \hline \hline
 & $\nu=-0.6$ & $-0.5$ & $-0.4$ & $-0.3$ \\ \hline
$m_H=115$ GeV & $>13.7$ TeV&$>13.9$ TeV  &$>14.8$ TeV &$>15.8$ TeV \\ \hline
     300 GeV   & $12.1-19.8$ TeV&$12.0-21.1$ TeV &$11.6-26.0$ TeV &$12.0-29.3$ 
TeV \\ \hline
    500 GeV    &$X$ &$X$ &$11.3-11.8$ TeV &$11.2-15.1$ TeV
 \\ \hline \hline
\end{tabular}
\caption{Table of $m_1$ ranges allowed by the EW precision data for several
representative values of $\nu$ and $m_H$.  An $X$ denotes that the parameter
choice corresponding to that location is not allowed.}
\label{m1range}
\end{table}

This relaxation of the upper bound on $m_H$ is
akin to that observed in~\cite{Rizzo:1999br}; the factors of 8.4 and 71 that
appear in the off-diagonal elements of the $W^{\pm}$ and $Z$ mixing matrices
here allow the effect to occur for much larger KK masses.  At this point the
reader may wonder whether these high $m_1$ values can be probed at future
colliders.  We will show in the next section that they are indeed 
invisible at the LHC; however, the large KK gauge boson couplings to 
third generation quarks produces observable effects over most of the allowed
parameter space at future $e^+ e^-$ colliders.

%\vspace*{-0.1cm}
\noindent
%\vspace*{0.1mm}

%\vspace*{-2.0cm}
\noindent

\section{Searches at the LHC}

\begin{figure}[htbp]
\centerline{
\psfig{figure=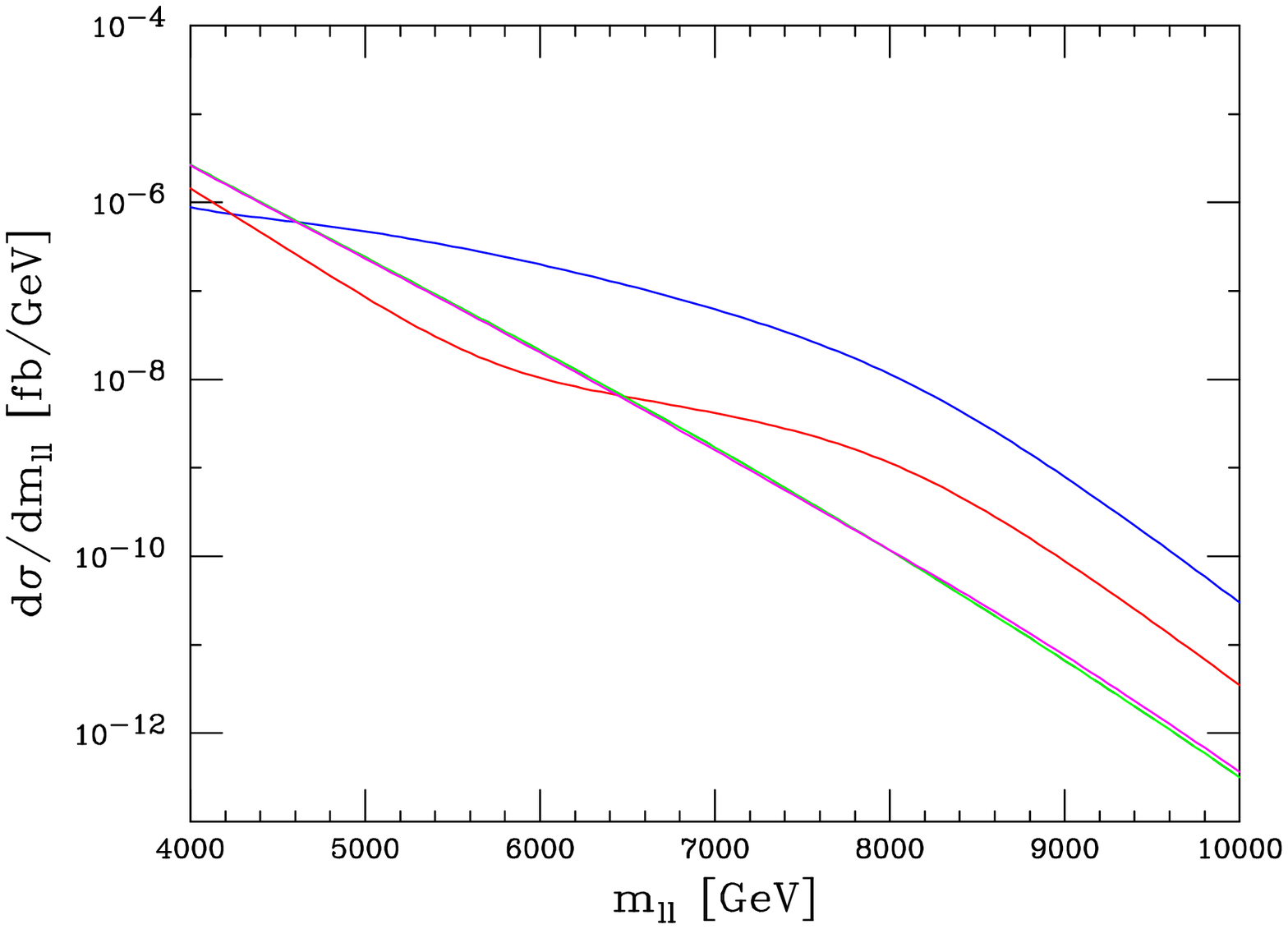,height=8.5cm,width=10.5cm,angle=0}}
\vspace{0.5cm}
\centerline{
\psfig{figure=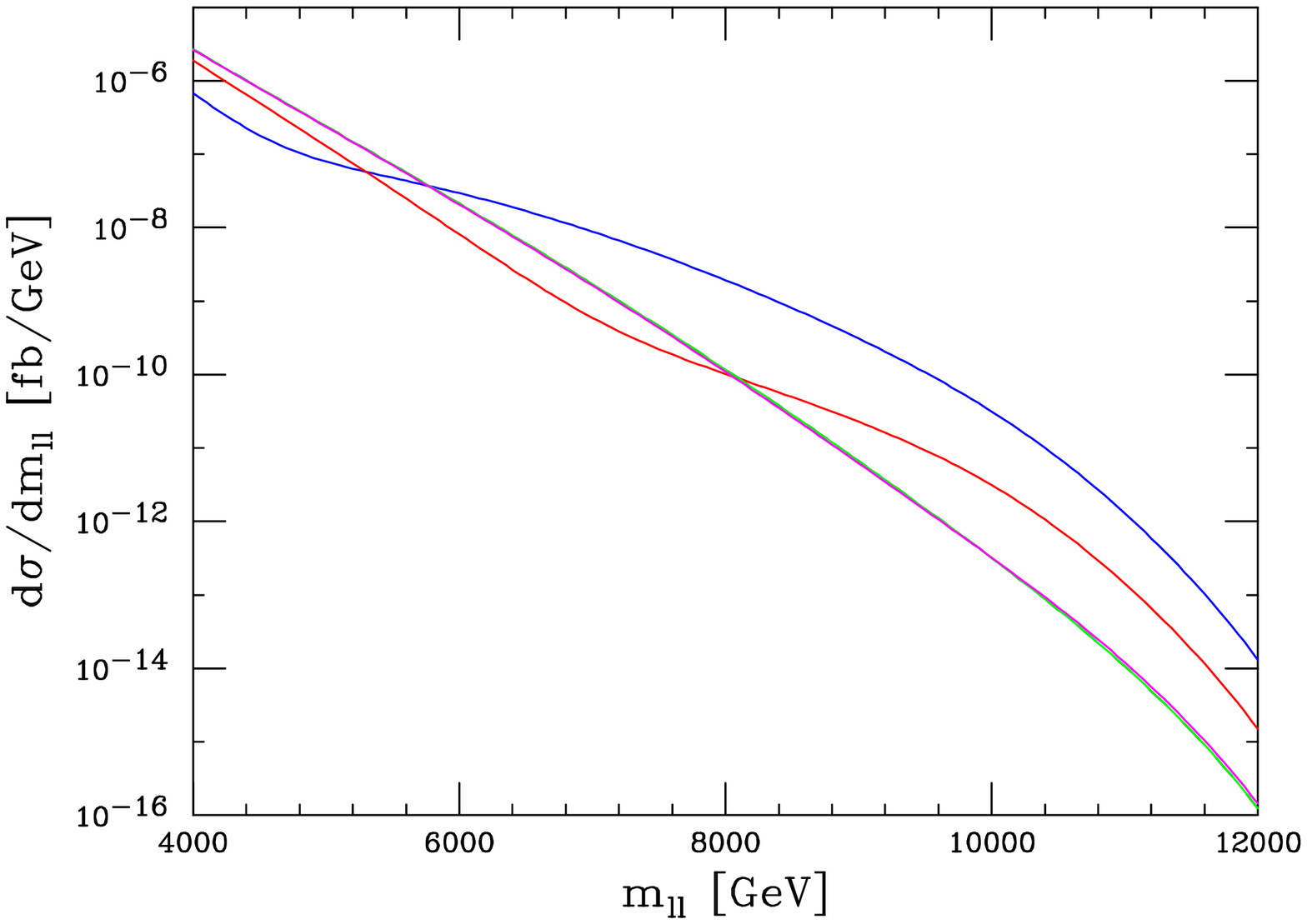,height=8.5cm,width=10.5cm,angle=0}}
\caption{Cross sections for the process $pp \rightarrow \mu^+ 
\mu^-$ for $m_1 = 8$ TeV (top) and $m_1 = 10$ TeV (bottom) as functions of 
the final state lepton pair invariant mass.  The upper blue curves are 
for $\nu =-0.3$, the slightly lower red curves represent
$\nu = -0.4$, and the three nearly degenerate straight lines correspond to
$\nu=-0.5, -0.6$, and the SM. K-factors and a rapidity cut
$|\eta| < 2.5$ have been included.}
\label{LHC}
\end{figure}

We now discuss the prospects for detecting the gauge KK states  which
are consistent with our EW fit at the LHC.  
The primary discovery mode for new heavy gauge bosons at hadron colliders is 
high invariant mass
Drell-Yan lepton pair production; at the LHC the relevant processes are 
$pp \rightarrow \gamma^{(1)},
Z^{(1)} \rightarrow \mu^+ \mu^-$, $e^+ e^-$.  The contributing parton 
level processes are $q\bar{q} \rightarrow \mu^+ \mu^-$, $e^+ e^-$.
We present $d\sigma/dm_{ll}$ for this process (with $m_{ll}$ being the
invariant mass of the final state lepton pair) in Fig.~\ref{LHC} for the 
parameter choices $\nu =-0.6,-0.5, -0.4, -0.3$ and $m_1 = 8,10$ TeV.  
These values are representative of the allowed region for $\nu$, but
the gauge KK masses are lighter than those allowed by the EW fit.
If the rates are 
unobservable at these points in parameter space, then the RS effects are 
undetectable for all interesting cases.  The resonances are wide in this
case primarily because of the large couplings of the 
KK gauge states to top and bottom quarks.  
 With the $100 \, 
{\rm fb}^{-1}$ of integrated luminosity envisioned for the LHC, the KK
contributions to Drell-Yan production are indeed invisible.  We present 
the expected number of excess events including both the $\mu^+ \mu^-$ and 
$e^+ e^-$ channels for this value of integrated luminosity and for the two 
choices of $\nu$ which produce the largest cross section 
in Table~\ref{LHCrates}.  Here, we have integrated over the   
invariant mass bins in which there is an excess of events over the SM 
predictions;
this corresponds to the cuts $m_{ll} \gsim 5$ TeV when $m_1 = 8$ TeV and 
$\nu = -0.3$, $m_{ll} \gsim 6.5$ TeV when $m_1 = 8$ TeV and $\nu = -0.4$,
$m_{ll} \gsim 6$ TeV when $m_1 = 10$ TeV and $\nu = -0.3$, and
$m_{ll} \gsim 8$ TeV when $m_1 = 10$ TeV and $\nu = -0.4$.  We have not 
attempted to study the depletion of events at lower $m_{ll}$ 
because the event rates at the affected invariant masses are too low. 
Two effects are hindering the detection of the KK contributions: the small 
couplings of zero mode fermions to KK gauge 
states for $\nu \leq -0.5$, and the high KK masses which require the parton 
subprocesses to occur at energies where the quark distribution functions 
are small.  Even with an order of magnitude increase in integrated luminosity,
the production of the first gauge KK excitation that is consistent
with the EW precision data is unobservable. 
\vspace{0.5cm}
\begin{table}[h]
\centering
\begin{tabular}{|r|r|r|} \hline \hline
 & $\nu=-0.4$ & $-0.3$ \\ \hline
$m_1 = 8$ TeV & $6.4 \times 10^{-4}$ & $8.8 \times 10^{-2}$ \\
$10$ TeV & $4.9 \times 10^{-6}$ & $3.2 \times 10^{-3}$  \\ \hline \hline
\end{tabular}
\caption{Table of expected Drell-Yan events at the LHC for various parameter
choices $L=100 \, {\rm fb}^{-1}$.  Both the $\mu^+ \mu^-$ and 
$e^+ e^-$ channels have been included.}
\label{LHCrates}
\end{table}

\begin{figure}[htbp]
\centerline{
\psfig{figure=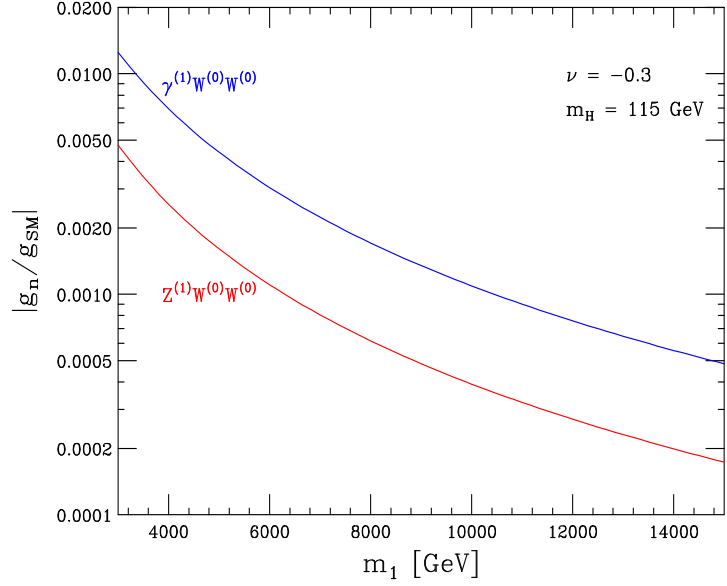,height=7.7cm,width=9.5cm,angle=0}}
\caption{Couplings of $W^{\pm,{(0)}}$ to $\gamma^{(1)}$
and $Z^{(1)}$, normalized to the SM couplings of $W^{\pm}$ to $\gamma$ and $Z$,
as functions of $m_1$.  The parameter values $\nu=-0.3$ and $m_H=115$ GeV have
been assumed.}
\label{TGCs}
\end{figure}

Another possible production mechanism for the KK gauge bosons at the LHC
is $W^+ W^-$ fusion, $pp\to WW+2$ jets$\to V^{(1)}+2$ jets.
The relevant triple gauge
couplings, $W^{+(0)}W^{-(0)}\gamma^{(1)}$ and $W^{+(0)}W^{-(0)}Z^{(1)}$,
are induced by mixing effects.  We present the strengths of these vertices 
normalized to the SM couplings $W^+ W^- \gamma$ and $W^+ W^- Z$ 
in Fig.~\ref{TGCs}.  Very slight 
$\nu$ and $m_H$ dependences enter these vertices; here we fix $\nu=-0.3$
and $m_H=115$ GeV, which maximizes their strength. For $m_1 \geq 11$ TeV,
these couplings are
a fraction, $\leq 10^{-3}$, of their SM strengths.  This, and the fact that the
$W^+ W^-$ fusion process is higher order in the EW coupling constant, render 
this a poor place in which to search for KK effects.

\begin{figure}[htbp]
\centerline{
\psfig{figure=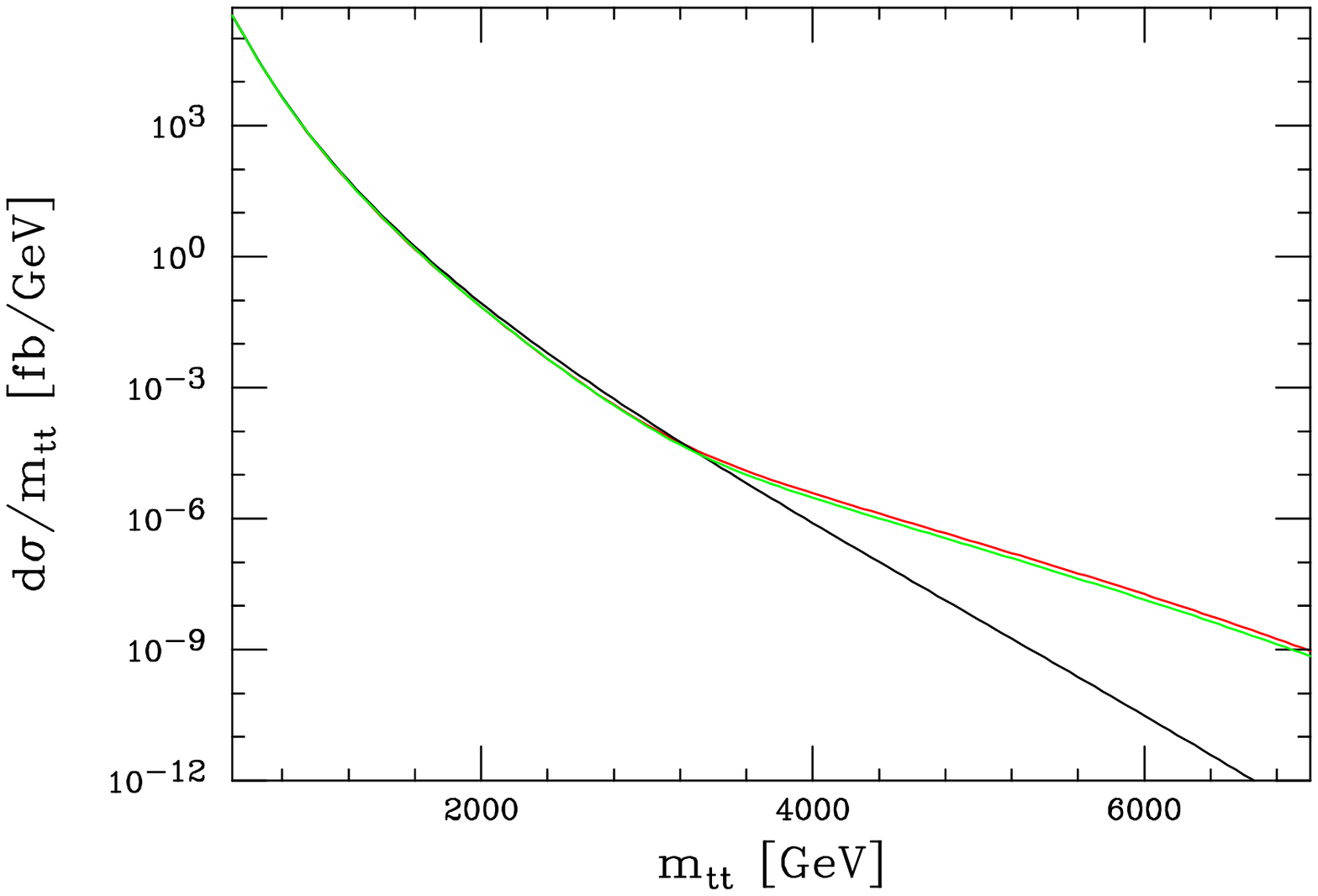,height=7.7cm,width=9.5cm,angle=0}}
\caption{Invariant mass distributions for $pp \rightarrow t\bar{t}$ at the
LHC including only the first KK state (top red line) and summing the first
five KK gauge bosons (bottom green line), for $m_1=10$ TeV.  The black line 
is the SM prediction.}
\label{topprod}
\end{figure}

The only remaining possibility for detecting the KK states at the LHC is via
deviations in top and bottom quark production.  These processes are
dominated at high energies by gluon initiated interactions; however,
these do not receive any modifications from gluon KK states since  
$g^{(0)}g^{(0)}g^{(n)}$ couplings do not exist.  We thus only examine top 
quark production, which receives a larger
contribution from quark initiated processes, and where the large couplings of
the KK gauge states to third generation quarks enter.  The invariant mass 
distribution is presented in
Fig.~\ref{topprod} for $m_1 =10$ TeV and $\nu=-0.4$.  Since the KK couplings 
to wall fermions do not decrease
with KK level, we have checked that the contributions from including
multiple states in the KK tower does not
significantly enhance the effect.  In fact, summing the first five KK 
contributions slightly decreases the cross section from that where only
the first level is included
due to the factor of $(-1)^n$ that enters the coupling of the $n$th KK level 
to top quarks.  The
expected number of excess events at the LHC is $\approx 0.14$, assuming a cut 
on the invariant mass of the final state top quarks of $m_{tt} \gsim 3.5$ TeV.
As in the previous case of Drell-Yan production, this event rate 
is undetectable even with an order of magnitude increase in integrated 
luminosity.  The slight depletion of events at lower invariant masses is 
similarly unobservable.  We must therefore conclude that the KK excitations 
which relax the precision EW upper bound on the Higgs mass are invisible at 
the LHC.
%
%\vspace*{-2.0cm}
\noindent
%
%\vspace*{-2.0cm}
\noindent
%
%\vspace*{-2.0cm}
\noindent

\section{Searches at Future TeV-scale Linear Colliders}

\begin{figure}[htbp]
\centerline{
\psfig{figure=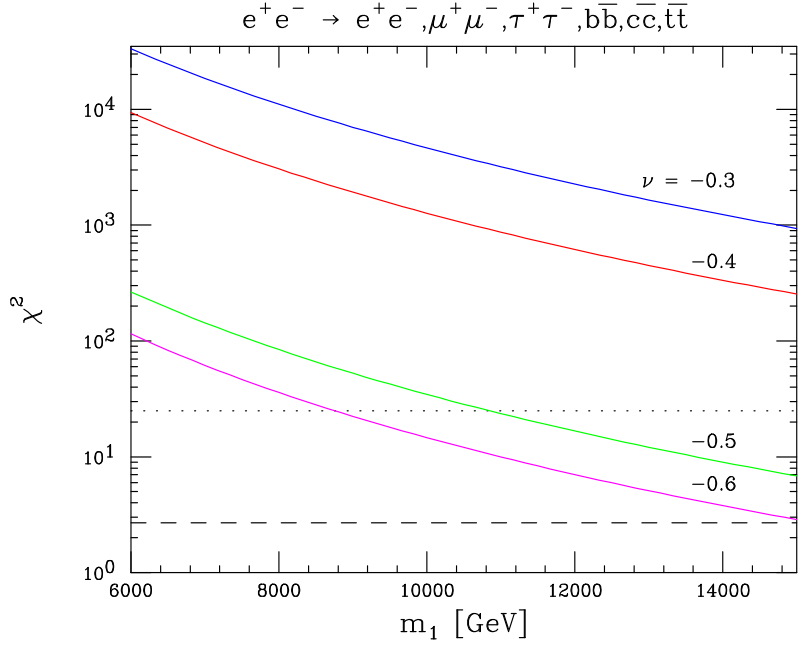,height=8.5cm,width=10.5cm,angle=0}}
\vspace{0.5cm}
\centerline{
\psfig{figure=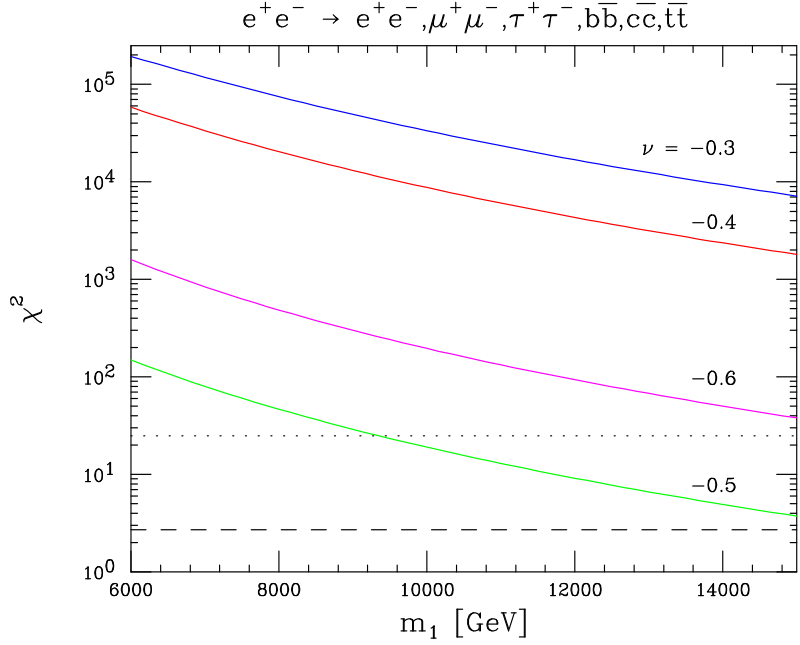,height=8.5cm,width=10.5cm,angle=0}}
\caption{$\chi^2$ values obtained by fitting the RS model predictions for
fermion pair production to the SM for $\nu=-0.6,-0.5,-0.4,-0.3$, as 
functions of $m_1$.  The upper figure assumes $\sqrt{s}=500$ GeV and
$L=500 \, {\rm fb}^{-1}$, while the lower assumes $\sqrt{s}=1000$ GeV and
$L=1000 \, {\rm fb}^{-1}$.  The dashed line indicates the $\chi^2$ 
necessary for exclusion of the model at the 95\% CL, and the dotted line
illustrates the $\chi^2$ required for a $5 \sigma$ discovery.  The 
polarizations and 
reconstruction efficiencies assumed are presented in the text.}
\label{NLCchi}
\end{figure}

We examine here whether KK gauge boson exchanges can be observed at a future 
$e^+ e^-$ collider with $\sqrt{s} =500-1000$ GeV and $L=500-1000 \, 
{\rm fb}^{-1}$.  Since the anticipated center-of-mass energies are well
below the 11 TeV KK gauge mass defining the lower edge of the allowed
range from the EW fit, 
we study the off-resonance modification of fermion pair production,
$e^+ e^- \rightarrow f\bar{f}$.  $Z$ pair production receives no KK gauge
contribution, while the
$\gamma^{(1)}$ and $Z^{(1)}$ exchanges in $e^+ e^- \rightarrow W^+ W^-$ suffer
from the weak triple gauge vertices displayed in Fig.~\ref{TGCs}.
We perform a $\chi^2$ fit to the
total rate, binned angular distribution, and binned ${\cal A}_{LR}$ for
fermion pair production to
estimate the search reaches possible at TeV-scale linear colliders.  We
assume an 80\% electron beam polarization, a $10^{\circ}$ angular cut, 
statistical errors and a 0.1\%
luminosity error.  We also use the following reconstructions efficiencies: 
a 100\% $\tau$ efficiency, a 70\% $b$ quark efficiency, a 50\%
$t$ quark efficiency, and a 40\% $c$ quark efficiency.  The $\chi^2$ values
obtained in this analysis are shown in Fig.~\ref{NLCchi} for several choices
of $\sqrt{s}$ and $L$.

We see from Fig.~\ref{NLCchi} that the effects of KK exchange exceed the
95\% CL exclusion limit for all $\nu$ values in the allowed region
and for $m_1 \leq 15$ 
TeV; the modifications when $\nu \geq -0.4$ reach the $5\sigma$ discovery 
limit.  The parameter space $\nu \leq -0.5$ and $m_1 > 15$ TeV, part of which 
provides a good fit to the EW precision data, falls below the exclusion limit.
It is possible that this difficulty can be alleviated with the inclusion
of more observables.  We note that a small hierarchy between 
the EW scale and $\Lambda_{\pi}$ begins to develop in this region, and it is 
consequently not as favored as the $m_1 \leq 15$ TeV range.
We note that the ordering of the $\nu = 
-0.6$ and $-0.5$ curves in the lower figure of Fig.~\ref{NLCchi} is correct.

%
%\vspace*{-2.0cm}
\noindent

\begin{figure}[htbp]
\centerline{
\psfig{figure=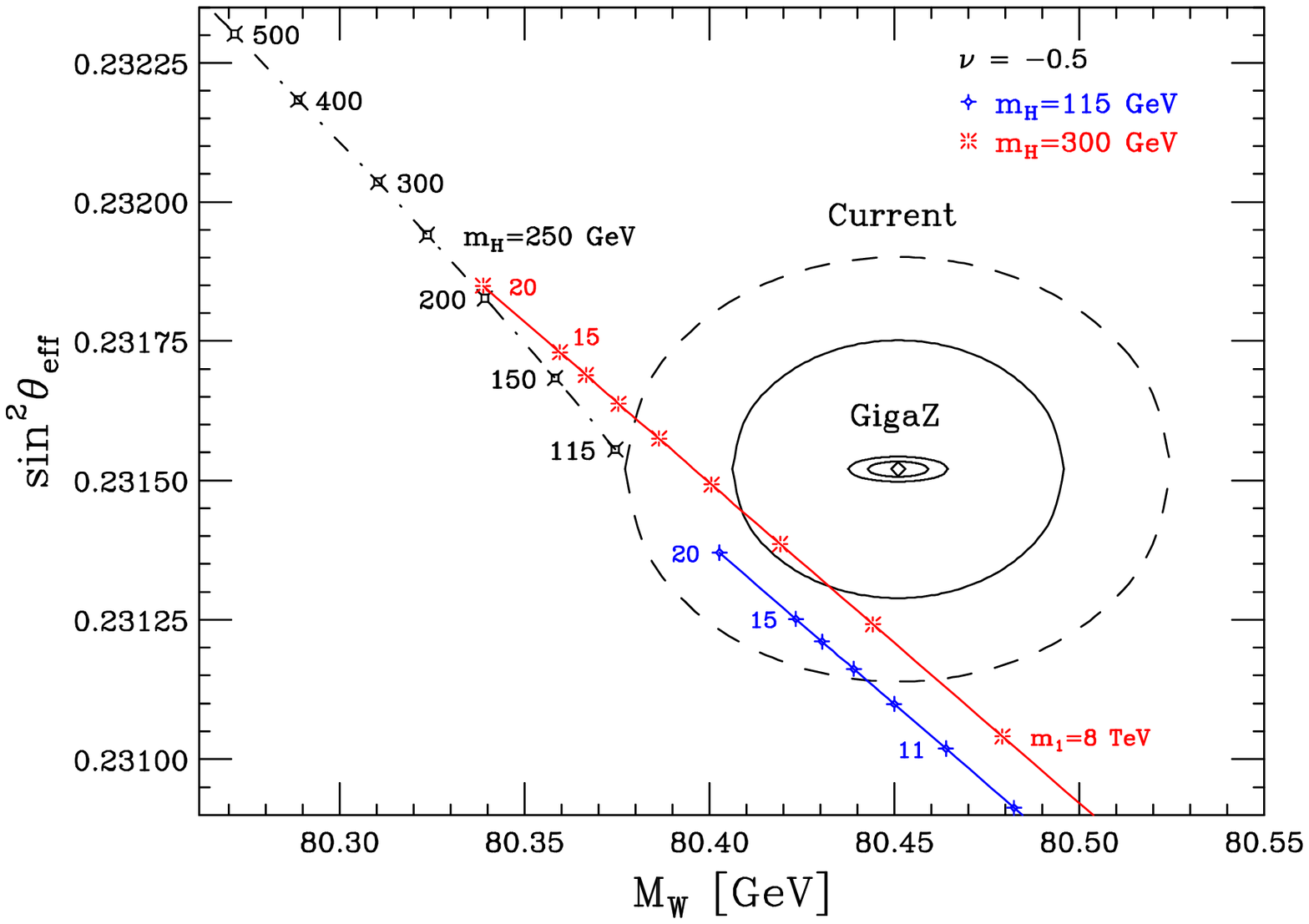,height=8.9cm,width=10.9cm,angle=0}}
\vspace{0.5cm}
\centerline{
\psfig{figure=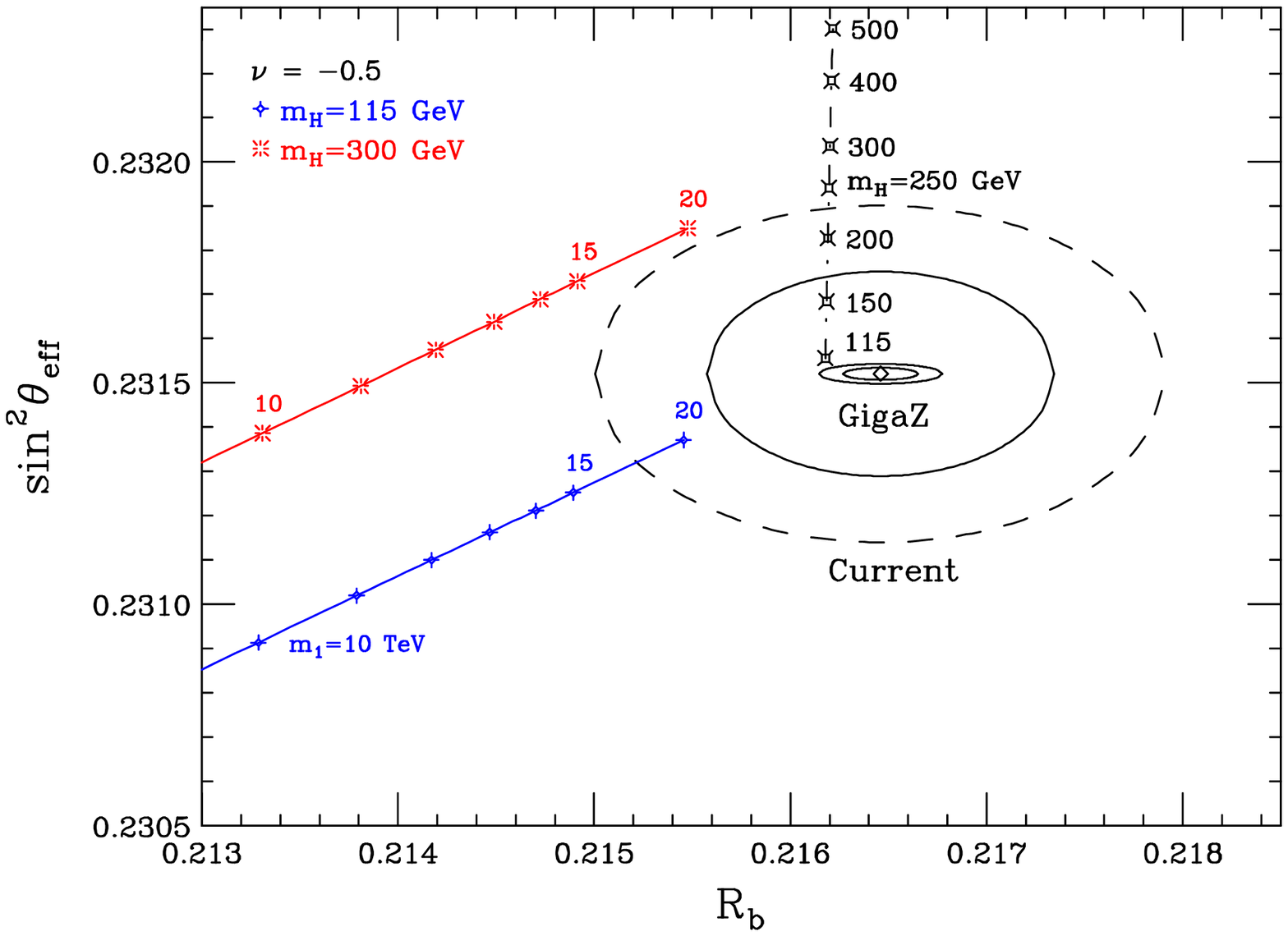,height=8.9cm,width=10.9cm,angle=0}}
\caption{The planes ${\rm sin}^{2}\theta_{eff}$ versus $M_W$ (top)
and ${\rm sin}^{2}\theta_{eff}$ versus $R_b$ (bottom) showing
current and future sensitivities, SM predictions, and RS model predictions.  
The diamonds show the current measured values of the observables.  The 
large solid and dashed ellipses represent respectively the 68\% and 95\% CL
regions from current sensitivities, while the smaller solid ellipses 
anticipate the same after operation of GigaZ.  The black dashdot lines show
the SM predictions for different Higgs
boson masses as labeled, while the solid colored lines show the RS model
results for varying $m_1$ for two Higgs masses satisfying the current 
EW constraints.}
\label{giga1}
\end{figure}

\begin{figure}[htbp]
\centerline{
\psfig{figure=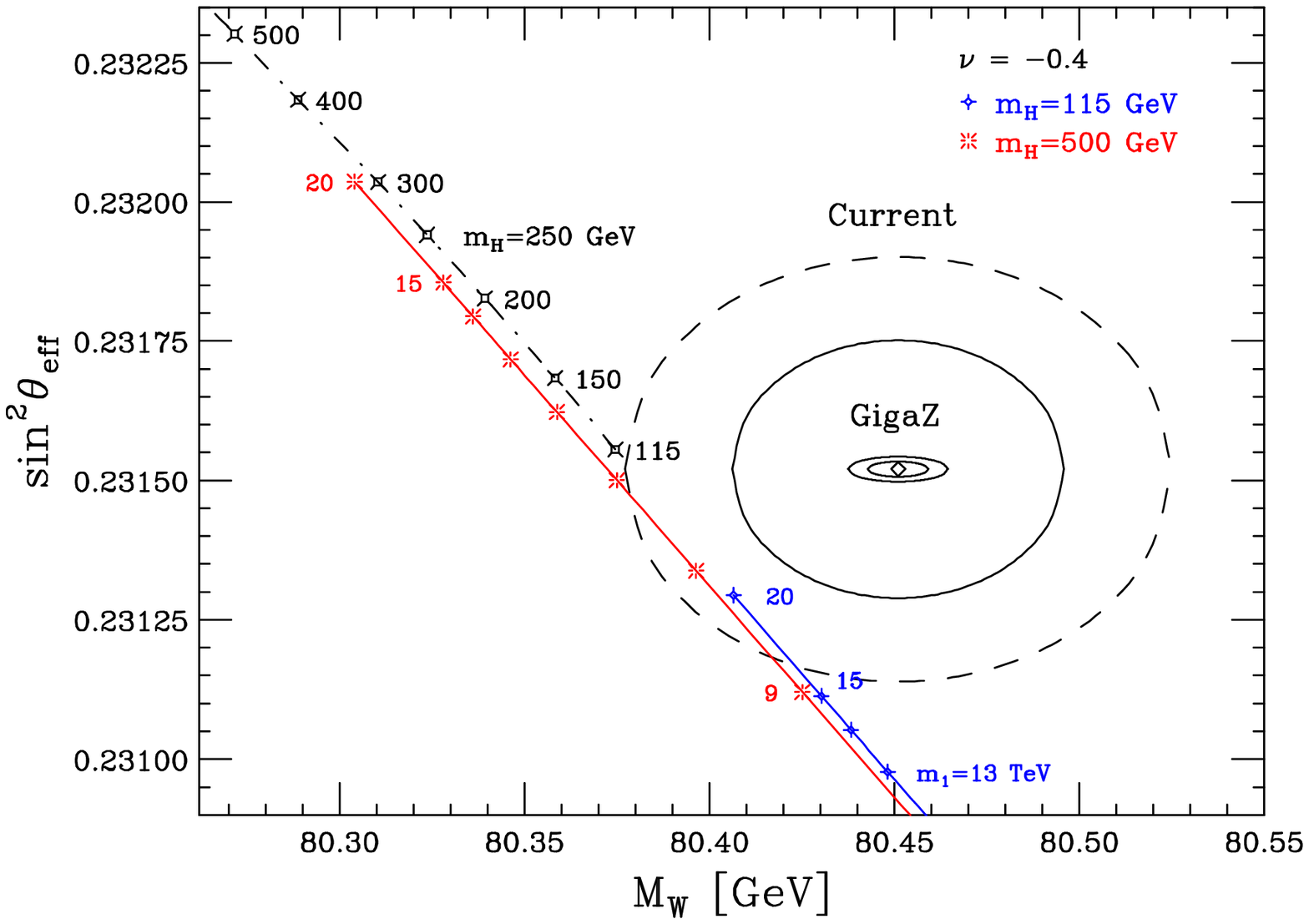,height=8.9cm,width=10.9cm,angle=0}}
\vspace{0.5cm}
\centerline{
\psfig{figure=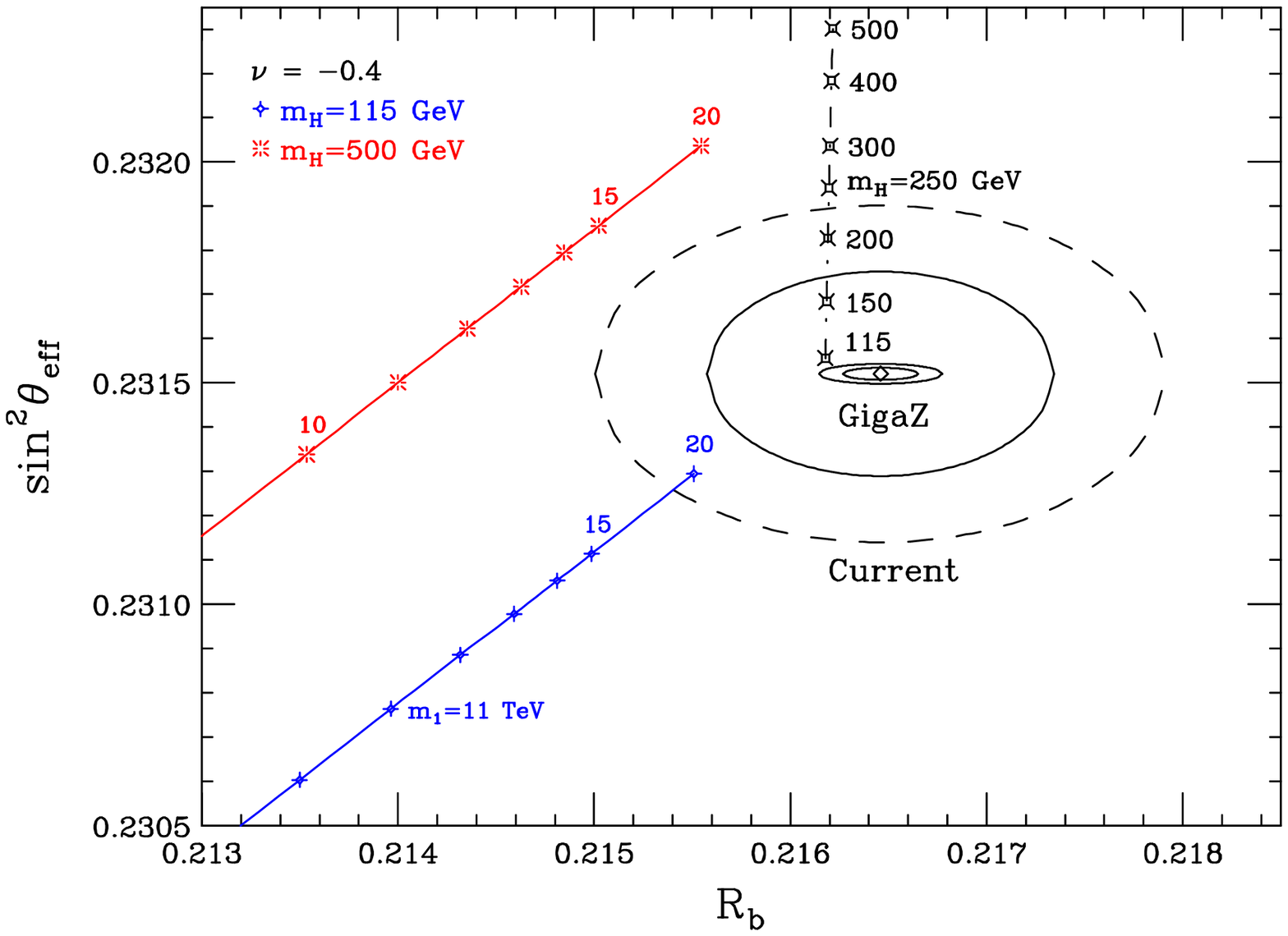,height=8.9cm,width=10.9cm,angle=0}}
\caption{Same as the previous figure for $\nu=-0.4$, and different $m_H$ 
choices.}
\label{giga2}
\end{figure}

We now subject this model to future high precision tests.  Planned $e^+ e^-$
colliders are designed for operation on the $Z$-pole for a period sufficient
to collect $10^9$ $Z$ events.  This program, known as GigaZ, will reduce
the error in ${\rm sin}^{2}\theta_{eff}$ to the $10^{-5}$ level and the
error in $R_b$ by a factor of 5~\cite{Aguilar-Saavedra:2001rg}.  A phase of 
operation on the $W^{\pm}$ pair production threshold is also planned,
which will reduce the error in the measurement of $M_W$ to 6 MeV.  
We now return to our analysis
of EW precision data and study the effects of this error reduction, 
keeping the central values for the observables unchanged from the
present and focusing on 
${\rm sin}^{2}\theta_{eff}$, $M_W$, and $R_b$.  Figures \ref{giga1} and
\ref{giga2} display our results in the
${\rm sin}^{2}\theta_{eff}$ versus $M_W$ plane and the 
${\rm sin}^{2}\theta_{eff}$ versus $R_b$ plane for $\nu=-0.5$ and $-0.4$.
These figures show the current and expected experimental precisions, 
SM predictions, and RS model results for
several different Higgs masses and $m_1$ choices.

%\vspace*{-1.0cm}
%
%\vspace*{-1.0cm}

It is difficult to predict what the status of fits to the EW precision data 
will be after the GigaZ program concludes, as a small shift in the 
experimental central values assisted by the small anticipated errors can 
drastically alter the current situation.  If the central values 
remain unchanged, it is clear from these figures that the improved precision
in the measurement of $R_b$ will disfavor the heavier Higgs 
solutions, and require a large value of $m_1$, which reintroduces a hierarchy
between $\Lambda_{\pi}$ and the EW scale.  However, the RS predictions for 
${\rm sin}^{2} \theta_{eff}$ and $M_{W}$ match experiment 
better than the SM results and can accommodate a heavy Higgs, and the global 
fit to the EW observables may prefer this solution.  Whatever scenario is 
realized, it is certainly true that the entire parameter space, including 
the region inaccessible in off-resonance fermion pair production, can be
probed at GigaZ.

\section{Constraints from FCNCs}

The placement of fermions in different locations in extra dimensions, on the 
TeV brane or in the bulk, leads to potentially dangerous FCNC since the 
Glashow-Weinberg-Paschos conditions~\cite{Glashow:1976nt,Paschos:1976ay} for 
the natural absence of FCNC are no longer satisfied.  These conditions
are violated automatically whenever fermions 
of different generations are treated asymmetrically by some form of new 
physics and mixing occurs between the relevant states. Within the RS scenario 
that we have constructed, these FCNC can arise from a number of potential 
sources, not all of which present the same level of danger.  A detailed 
analysis of FCNC effects is certainly beyond the scope of 
this paper and requires a specific flavor model as input; we simply outline 
the potential sources of FCNC and provide a few estimates of their size. 

The most obvious sources of FCNC are from the exchanges of gauge bosons.  The 
states in the gauge KK towers can feel the different fermion generation 
localities, and through intergenerational mixing can then induce FCNC. 
Furthermore, the couplings of the wall fields to the KK gauge states are 
enhanced by a factor of $\approx \sqrt {2\pi kr_c}$.  Since zero mode KK 
gauge states
in the limit of vanishing mixing are constrained by construction to have the 
same couplings to fermions as do the SM gauge bosons, such fields can only
induce FCNC through the small admixture of KK weak eigenstates introduced by 
mixing.  These effects are suppressed by small mixing angles, and are not
as important as those arising from the KK towers themselves.  
We therefore expect that
the KK gauge state contributions represent the greatest source of 
potentially dangerous FCNC.

Graviton KK towers can also probe the different locations 
of the SM fermion generations and induce FCNC-like couplings. However, in 
this case the potentially dangerous contributions are much smaller since 
($i$) graviton-induced FCNC take the form of dimension-8 operators, in 
contrast to the dimension-6 KK gauge contributions, and lead to amplitudes 
which are suppressed 
by factors of order $m_{K,D,B}^2/\Lambda_\pi^2$. This is an enormous degree of 
suppression since we have shown that $\Lambda_\pi \gsim 10$ TeV in the 
scenario presently under consideration. ($ii$) Unlike KK gauge fields, the 
graviton KK couplings to wall fields are not enhanced by the
factor $\sqrt {2\pi kr_c}$.

How large are the KK gauge tower contributions? The answer depends upon 
which gauge boson we are examining.  We neglect in this analysis the small
mixing between the first and second generation fermions and their KK towers.  
Let $g_{L,R}^a$ represent the couplings of a particular 
fermion with electric charge $Q$ to one of the neutral SM gauge bosons 
labeled by the index $a$.  We write the fermion couplings to KK gauge states
as $g_{L,R}^a c^n(\nu_i)$, where $\nu_i$ is the $i$th generation bulk mass
parameter and $n$ labels the gauge KK tower level.  Note that the functions
$c^n$ in the present model are independent of 
chirality and the gauge boson under consideration. The fact that the 
$c^n(\nu_i)$ are different for each $i$ generates the FCNC terms 
when we transform to the mass eigenstate basis. Let $U_{L,R}$ represent the 
matrices performing the bi-unitary transformation required to diagonalize the 
appropriate fermion mass matrix.  The off-diagonal couplings in the mass 
eigenstate basis are then given by 
\begin{equation}
(Q^n_{L,R})^a_{ij}=g_{L,R}^a \sum_k (U_{ik})_{L,R} c^n(\nu_k)(U^\dagger_{kj})
_{L,R}\,. 
\end{equation}
For the specific model discussed in the previous sections we have 
$c^n(\nu_1)=c^n(\nu_2)\neq c^n(\nu_3)$, and we use the unitarity of the 
$U$'s to rewrite these couplings as 
\begin{equation}
(Q^n_{L,R})^a_{ij}=g_{L,R}^a [c^n(\nu_3)-c^n(\nu_1)](U_{i3}U^\dagger_{3j})_
{L,R}\,. 
\end{equation}
With the third generation on the wall and the first and second in the bulk in 
the region $-0.6 \leq \nu_1 \leq -0.3$, it is clear that 
$|c^n(\nu_3)-c^n(\nu_1)| \simeq \sqrt {2\pi kr_c}$ for all $n$; {\it at 
worst}, the size of the off-diagonal couplings in our model is given by 
\begin{equation}
(Q^n_{L,R})^a_{ij}=\sqrt {2\pi kr_c}\, g_{L,R}^a (U_{i3}U^\dagger_{3j})_{L,R}
\,, 
\end{equation}
which is independent of $n$.  The $U_{ij}$ arise from some complete theory
of flavor that must reproduce the experimentally measured CKM matrix $V_{ij}$. 
We therefore expect $U_{ij} \simeq V_{ij}$ and ee adopt this approximation 
in our estimates below.

The most stringent constraints on FCNC arise from low energy processes such
as meson-antimeson mixing and rare decays~\cite{Groom:in}; we present here 
our estimate for $K-\bar{K}$ mixing. The above interaction generates a 
coupling which can be symbolically written as 
\begin{equation}
{\cal L}=2\pi kr_c\sum_a \sum_n (J^a_L+J^a_R)^2/m_n^2\,, 
\end{equation}
where $J^a_{L,R}=g_{L,r}^a V_{i3}V^\dagger_{3j} \bar f_i\gamma_\mu P_{L,R}
f_j$, $m_n$ is the mass of the $n$th KK gauge state, and we have summed all KK 
contributions.  Recalling the lore 
that we can accurately approximate the matrix element of the two currents in 
the vacuum insertion approximation, we see that the KK gluon 
towers do not contribute. This is due to the fact that these states only 
couple to currents with non-zero color while both the meson and the vacuum 
are color singlets. Thus we need to consider only the $Z$ and $\gamma$ tower 
exchanges.  Using 
$\sum_n m_n^{-2}\simeq 1.5m_1^{-2}$~\cite{Davoudiasl:1999tf}, 
$V_{13}V^\dagger_{32}\simeq A^2 (1-\rho)^2 \lambda^5$ in the Wolfenstein 
parameterization and 
\begin{equation}
<K|J_LJ_R|\bar K>=\Omega_K <K|J_LJ_L|\bar K>=\Omega_K <K|J_RJ_R|\bar K>\,,
\end{equation} 
with $\Omega_K \simeq 7$~\cite{Beall:1981ze} for the current-current matrix 
element, we arrive at 
\begin{equation}
{{\Delta m_{KK}^{RS}}\over {\Delta m_{KK}^{SM}}}\simeq 0.0098[1+0.73\Omega_K] 
\left(11 TeV\over {m_1} \right)^2 \simeq 0.06\,,
\end{equation}
which is within the uncertainty of the SM result~\cite{Buras:2001pn}. 
From this estimate we see 
that, at least for the $K-\bar K$ system, the RS FCNC contributions are 
rather small.  We have also studied $B-\bar{B}$ mixing and obtain 
similar results.

Once a realistic theory of flavor within this RS model context is 
constructed, we can perform a more detailed and quantitative analysis of the 
potential impact of FCNC.  It will be interesting to examine if existing 
bounds can provide further constraints on the RS model parameters 
within such a framework.

\section{Discussions and Conclusions}

In this paper we have re-examined the placement of SM fermions in the full 
5-dimensional bulk of the Randall-Sundrum spacetime.  We have found that 
mixing between the top quark zero mode and its KK tower, induced by the large 
top quark mass, yields shifts in the $\rho$ parameter that are inconsistent 
with current measurements.  To obviate these bounds we must take the 
fundamental RS scale $\Lambda_{\pi} \gsim 100$ TeV, reintroducing the 
hierarchy between the Planck and EW scales and thus destroying the original 
motivation for the RS model.  We instead proposed a mixed scenario which 
localizes the third generation of quarks, and presumably leptons, on the 
TeV-brane and allows the lighter two generations to propagate in the RS bulk.  
For values of the bulk mass parameter in the region $-0.55 \lsim \nu 
\lsim -0.35$, the same values allowed by both contact interaction searches 
and $\rho$ parameter constraints arising from the first two generations, the 
fermions mass hierarchies $m_c / m_t$ and $m_s / m_b$ are naturally reproduced.

We next explored the consequences of this proposal for current precision EW 
measurements.  We studied modifications of the electroweak observables caused 
by both mixing of the SM gauge bosons with their corresponding KK towers and 
the exchanges of higher KK states; we found that with KK masses $m_1 
\approx 11$ TeV and bulk mass parameters $\nu \approx -0.5,-0.4$ a Higgs 
boson with mass $m_H \lsim 500$ GeV can provide a good fit to the precision 
electroweak data.  An analysis of the fit showed that the large couplings 
between the zero mode bottom quark and KK gauge bosons induced large shifts 
in $R_b$ that prevented a heavier Higgs from being consistent with the 
precision data.

We then examined the signatures of this scenario at future high energy 
colliders.  We found that the parameter region consistent with the 
precision electroweak data does not lead to any new physics signatures at the 
LHC; the expected event excess in both Drell-Yan and gauge boson fusion 
processes are statistically insignificant with the envisioned integrated 
luminosities, and the predicted modification of the $t\bar{t}$ production 
cross section is similarly unobservably small.  The only new physics that the 
LHC would possibly observe is a Higgs boson apparently heavier than that 
allowed by the SM electroweak fits.  By contrast, the parameter range 
$m_1 \lsim 15$ 
TeV and $\nu \lsim -0.3$ can be probed in fermion pair production processes 
at a future $e^+ e^-$ collider with center-of-mass energy of $500-1000$ GeV, 
while the region $m_1 \lsim 25-30$ TeV and $\nu \lsim -0.3$ is testable at 
GigaZ.  For larger KK first excitation masses, we reintroduce the 
hierarchy between $\Lambda_{\pi}$ and the electroweak scale.

Finally, we considered the possible constraints on this scenario arising from 
low energy FCNC. The asymmetric treatment of the three fermion generations 
allows KK $Z$-boson exchanges to mediate FCNC interactions.  We estimated the 
contributions of such effects to meson-antimeson mixing, and found that 
their size is within the theoretical errors inherent to meson mixing.
However, a detailed analysis of 
FCNC effects requires a full model of flavor, which we have not constructed.

In summary, we have found that the experimental restrictions on placing SM 
matter in the RS bulk lead naturally to a very interesting region of parameter 
space.  This parameter region provides a geometrical origin for the fermion 
Yukawa hierarchies, and allows a heavy Higgs boson to be consistent with 
precision measurements while remaining otherwise invisible at the LHC.  We 
believe that such features render this model worthy of further study.

\bigskip

\noindent
{\Large \bf Acknowledgements}

\noindent
F.P. would like to thank K. Melnikov for discussions.  The work of F.P. was 
supported in part by the National Science Foundation 
Graduate Research Program.

%
%%%%%%%%%%%%%%%%%%--- References
%%%%%%%%%%%%%%%%%%%%%%%%%%%%%%%%%%%%%%%%%%%%%%%%%%%%%%%
\def\MPL #1 #2 #3 {Mod. Phys. Lett. {\bf#1},\ #2 (#3)}
\def\NPB #1 #2 #3 {Nucl. Phys. {\bf#1},\ #2 (#3)}
\def\PLB #1 #2 #3 {Phys. Lett. {\bf#1},\ #2 (#3)}
\def\PR #1 #2 #3 {Phys. Rep. {\bf#1},\ #2 (#3)}
\def\PRD #1 #2 #3 {Phys. Rev. {\bf#1},\ #2 (#3)}
\def\PRL #1 #2 #3 {Phys. Rev. Lett. {\bf#1},\ #2 (#3)}
\def\RMP #1 #2 #3 {Rev. Mod. Phys. {\bf#1},\ #2 (#3)}
\def\NIM #1 #2 #3 {Nuc. Inst. Meth. {\bf#1},\ #2 (#3)}
\def\ZPC #1 #2 #3 {Z. Phys. {\bf#1},\ #2 (#3)}
\def\EJPC #1 #2 #3 {E. Phys. J. {\bf#1},\ #2 (#3)}
\def\IJMP #1 #2 #3 {Int. J. Mod. Phys. {\bf#1},\ #2 (#3)}
\def\JHEP #1 #2 #3 {J. High En. Phys. {\bf#1},\ #2 (#3)}

\end{document}